\documentclass[aps,twocolumn,amsmath]{revtex4}
\usepackage{graphicx}
\usepackage{dcolumn}
\def\Journal#1#2#3#4{{#1} {\bf#2}, #3 (#4)}

\def\PLB{{\rm Phys. Lett.}  B}
\def\PRL{\rm Phys. Rev. Lett.}
\def\PRD{{\rm Phys. Rev.} D}

\def\JPG{{\rm J. Phys.} G}
\def\up{\uparrow}
\def\down{\downarrow}

\def\ep{\epsilon}
\def\lam{\lambda}

\def\la{\langle}
\def\ra{\rangle}

\def\al{\alpha}

\def\be{\begin{equation}}
\def\ee{\end{equation}}
\def\bea{\begin{eqnarray}}
\def\eea{\end{eqnarray}}

\begin{document}
\title{Perturbative QCD analysis of exclusive
$J/\psi+\eta_c$ production in $e^+e^-$ annihilation}
\author{ Ho-Meoyng Choi$^{a}$ and Chueng-Ryong Ji$^{b}$\\
$^a$ Department of Physics, Teachers College, Kyungpook National University,
     Daegu, Korea 702-701\\
$^b$ Department of Physics, North Carolina State University,
Raleigh, NC 27695-8202, USA}
\begin{abstract}
We analyze the exclusive charmonium $J/\psi+\eta_c$ pair production in
$e^+e^-$ annihilation using the nonfactorized perturbative QCD
and the light-front quark model(LFQM) that goes beyond the peaking 
approximation. We effectively include all orders of higher twist terms
in the leading order of QCD coupling constant and compare our
nonfactorized analysis with the usual factorized analysis in the calculation
of the cross section.
We also calculate the quark distribution amplitudes, the Gegenbauer moments,
and the decay constants for $J/\psi$ and $\eta_c$ mesons using our LFQM. 
Our nonfactorized result enhances the NRQCD result
by a factor of $3\sim4$ at $\sqrt{s}=10.6$ GeV.
\end{abstract}


\maketitle
\section{Introduction}
It has been known that
the exclusive pair production of heavy meson can be reliably predicted 
within the framework of perturbative 
quantum chromodynamics(PQCD), since the wave function is well
constrained by the nonrelativistic consideration~\cite{BJ85}. 
However, the large discrepancy between the theoretical 
predictions~\cite{BJ03,LHC03,HKQ03,Kis} based on the
nonrelativistic QCD(NRQCD)~\cite{VS} factorization approach
and the experimental results~\cite{Belle,Babar} for the exclusive
$J/\psi+\eta_c$ production in $e^+e^-$ annihilation at the energy
$\sqrt{s}=10.6$ GeV has
triggered the need of better understanding both in the available
calculational tools and the appreciable relativistic effects.

A particularly 
convenient and intuitive framework in applying PQCD to exclusive
processes is based upon the light-front(LF) Fock-state decomposition 
of hadronic state.
If the PQCD factorization theorem is applicable, then the invariant 
amplitude
for exclusive processes factorizes into the convolution of the valence
quark distribution amplitudes(DAs) $\phi(x,q^2)$ with the hard scattering
amplitude $T_H$, which is dominated by one-gluon
exchange diagrams at leading order of QCD coupling constant $\al_s$. 
To implement the factorization theorem at high momentum transfer, the 
hadronic wave function plays an important role linking between the long
distance nonperturbative QCD and the short distance PQCD. 
In the LF framework, the valence quark DA is computed from the valence
LF wave function
$\Psi_n(x_i,{\bf k}_{\perp i})$
of the hadron at equal LF time $\tau=t + z/c$ which
is the probability amplitude to find $n$
constituents(quarks,antiquarks, and gluons) with LF momenta
$k_i=(x_i,{\bf k}_{\perp i})$ in a hadron. Here, $x_i$ and ${\bf k}_{\perp i}$
are the LF longitudinal momentum fraction and the transverse momenta of the $i$th
constituent in the $n$-particle Fock-state, respectively.

The NRQCD factorization approach~\cite{BJ03,LHC03,HKQ03,Kis} for charmonium
production assumes that the constituents are sufficiently nonrelativistic
so that the relative motion of valence quarks can be neglected inside the 
meson. In this case, the quark DA becomes the $\delta$ function, i.e. 
$\phi(x,q^2)\sim\delta(x-1/2)$(the so-called peaking approximation). 
However, the cross section value~\cite{BJ03,LHC03,HKQ03,Kis} 
estimated within the NRQCD factorization approach in the leading order 
of $\al_s$ underestimates the experimental 
data~\cite{Belle,Babar} by an order of magnitude. 
In order to reduce the discrepancy between theory and experiment, the
authors in Refs.~\cite{BC,Ma,BLL05,Huang} considered 
a rather broad quark DA instead of $\delta$-shaped quark DA. 

However, as pointed out in Refs.~\cite{JP,CJD}, if the quark DA is not an
exact $\delta$ function, i.e. ${\bf k}_\perp$ in the soft bound state 
LF wave function can play a significant role, the factorization theorem 
is no longer applicable. To go beyond the peaking approximation, the 
invariant amplitude should be expressed in terms of the LF wave function
$\Psi(x_i,{\bf k}_{\perp i})$ rather than the quark DA.
In Refs.~\cite{JP,CJD}, we discussed the validity issue of peaking 
approximation for the heavy pseudoscalar meson pair production 
processes such as $e^+e^-\to P+P$($P=B_c,B_s,B,D,D_s$) 
using the LF model wave function $\Psi(x_i,{\bf k}_{\perp i})
\propto \exp(-M^2_0/\beta^2)$,
where $M_0$ is the invariant mass of the constituent quark and antiquark
defined by $M^2_0=\sum_i({\bf k}^2_\perp + m^2_i)/x_i$ and
$\beta$ is the gaussian parameter.
The gaussian parameter $\beta$ in our model wave function was 
found to be related to the transverse momentum via
$\beta=\sqrt{\la{\bf k}^2_\perp\ra}$. This relation
naturally explains the zero-binding energy limit as the zero transverse
momentum, i.e.  $\la M^2_0\ra= (m_1 + m_2)^2$ and $x_i=m_i/M$
for $\beta= 0$. We also found that the heavy quark DA is sensitive 
to the value of $\beta$ and indeed quite different from the $\delta$-type DA
according to our LFQM based on the variational principle for
the QCD-motivated Hamiltonian~\cite{CJ1,CJ2}.
In going beyond the peaking approximation, we stressed a consistency
of the formulation
by keeping the transverse momentum ${\bf k}_\perp$ both in the 
wave function part and the hard scattering part together before doing any 
integration in the amplitude. 
Similar consideration has also been made in the recent investigation
of the relativistic and bound state effects\cite{EM} not based on the 
light-front dynamics(LFD) but
including the relativistic effects up to the second order of the
relative quark velocity, i.e. $\la v^2\ra$.
Such non-factorized analysis should be 
distinguished from the factorized analysis~\cite{BC,Ma,BLL05} 
where the transverse momenta are seperately integrated out in the
wave function part and in the hard scattering part.
Even if the used LF wave functions lead to the similar shapes of DAs,
it is apparent that the
predictions for the cross sections of heavy meson productions would
be different between the factorized and non-factorized analyses.

In this work, we extend our previous works~\cite{JP,CJD} of pseudoscalar
meson pair production to 
the case of pseudoscalar and vector meson productions and calculate
the cross section for $e^+e^-\to J/\psi+\eta_c$ process at leading order
of $\al_s$ including effectively all orders of higher twist terms. 
As noted in~\cite{CJD}, our results for the quark DA
of $J/\psi$ and $\eta_c$ are quite different from the $\delta$-type 
function. We find that the non-factorized form of the form factor 
enhances the cross section of NRQCD result by a factor of $3\sim4$ 
at $\sqrt{s}=10.6$ GeV while 
it reduces that of the factorized formulation by 20$\%$.
Since the cross section for $e^+e^-\to J/\psi+\eta_c$
is found to be very sensitive to the behavior of the end points
($x\to 0$ and 1) in the quark DA, we also examine the  
results of the decay constants or equivalently the Gegenbauer moments 
of $J/\psi$ and $\eta_c$ mesons. Since the perturbative corrections of order $\al_s$
to the production amplitude has already been obtained\cite{ZGC} increasing
the cross section significantly,
it is important to consider the more accurate
assessment of cross section at the leading order of $\al_s$.

The paper is organized as follows.
In Sec. II, we describe the formulation of our
light-front quark model (LFQM), which has been quite
successful in describing the static and non-static properties of the
pseudoscalar and vector mesons~\cite{CJ1,CJ2}. The formulae for the quark DA,
decay constants, Gegenbauer and $\xi(=x_1-x_2)$ moments are 
also given in this section. In Sec. III, the transverse momentum dependent
hard scattering amplitude and the form factor for $e^+e^-\to J/\psi+\eta_c$ 
transition are given in leading order of $\al_s$.
The form factors both in the factorized and nonfactorized  
formulations are
explicitly given in this section. We also show in this section that 
our peaking approximation(i.e. NRQCD) result coincides 
with the one derived from Ma and Si~\cite{Ma}.
In Sec.IV, we present the numerical results for the decay constants, 
quark DAs, Gegenbauer and $\xi$ moments for the $J/\psi$ and $\eta_c$ mesons
and compare them with other theoretical model predictions in addition to the available
experimental data. The numerical results for the
$e^+e^-\to J/\psi + \eta_c$ cross section are obtained and compared
with the data~\cite{Belle,Babar}. Summary and conclusions follow in Sec. V.
In the Appendices A and B, we summarize our results for the helicity 
contributions to the hard scattering amplitudes and the form factor,
respectively.

\section{Model Description}
In our LFQM~\cite{CJ1,CJ2}, 
the momentum space light-front wave function of the ground state
pseudoscalar and vector mesons is given by
\be\label{w.f}
\Psi^{JJ_z}_{100}(x_i,{\bf k}_{i\perp},\lam_i)
={\cal R}^{JJ_z}_{\lam_1\lam_2}(x_i,{\bf k}_{i\perp})
\phi_R(x_i,{\bf k}_{i\perp}),
\ee
where $\phi_R(x_i,{\bf k}_{i\perp})$ is the radial wave function and
${\cal R}^{JJ_z}_{\lam_1\lam_2}$ is the spin-orbit wave function
obtained by the interaction independent Melosh transformation
from the ordinary equal-time static spin-orbit wave function assigned
by the quantum numbers $J^{PC}$.
The model wave function in Eq.~(\ref{w.f}) is represented by the
Lorentz-invariant variables, $x_i=p^+_i/P^+$,
${\bf k}_{i\perp}={\bf p}_{i\perp}-x_i{\bf P}_\perp$ and $\lam_i$, where
$p^\mu_i$ and $\lam_i$ are the momenta and the helicities of
constituent quarks, respectively, and $P^\mu=(P^+,P^-,{\bf P}_\perp)
=(P^0+P^3,(M^2+{\bf P}^2_\perp)/P^+,{\bf P}_\perp)$ is the momentum of the
meson $M$.

The covariant forms of the spin-orbit wave functions
for pseudoscalar and vector mesons are respectively given by
\bea\label{R00_A}
{\cal R}_{\lam_1\lam_2}^{00}
&=&\frac{-\bar{u}(p_1,\lam_1)\gamma_5 v(p_2,\lam_2)}
{\sqrt{2}M_0},
\nonumber\\
{\cal R}_{\lam_1\lam_2}^{1J_3}
&=&\frac{-\bar{u}(p_1,\lam_1)
\biggl[/\!\!\!\ep(J_z) -\frac{\ep\cdot(p_1-p_2)}{M_0 + 2m}\biggr]
v(p_2,\lam_2)} {\sqrt{2}M_0},
\nonumber\\
\eea
where $\ep^\mu(J_z)$ is the polarization vectors of the vector meson,
$M^2_0=({\bf k}^2_{\perp}+m^2)/x_1x_2$ is the invariant meson 
mass square, and $\sum_{\lam_1\lam_2}{\cal R}_{\lam_1\lam_2}^{JJ_z\dagger}
{\cal R}_{\lam_1\lam_2}^{JJ_z}=1$ for both pseudoscalar and vector mesons.
Using the four-vectors $p_1,p_2$
given in terms of the LF relative momentum variables
$(x,{\bf k}_{\perp})$ as 
\bea
&&p^+_1= x_1P^+,\;\; p^+_2=x_2P^+, \nonumber\\
&&{\bf p}_{1\perp} = x_1{\bf P}_\perp + {\bf k}_{\perp},\;
\; {\bf p}_{2\perp}=x_2{\bf P}_{\perp} - {\bf k}_{\perp},
\eea
we obtain the explicit forms of spin-orbit wave functions for pseudoscalar
and vector mesons with the longitudinal($\ep(0)$) and
transverse($\ep(+1)$) polarizations as follows 
\be\label{R00}
{\cal R}^{00}_{\lam_1\lam_2}\:=
\frac{1}{C}
\left(
\begin{array}{cc}
        -k^L & m\\
        -m & -k^R
      \end{array}
    \right),\;
\ee
\be\label{R10}
{\cal R}^{10}_{\lam_1\lam_2}\:=
\frac{1}{C}
\left(
\begin{array}{cc}
k^L\frac{(1-2x)M_0}{M_0+2m} &
m + \frac{2{\bf k}^2_\perp}{M_0 + 2m}\\
m + \frac{2{\bf k}^2_\perp}{M_0 + 2m} &
-k^R\frac{(1-2x)M_0}{M_0+2m}
      \end{array}
    \right),\;
\ee
\be\label{R11}
{\cal R}^{11}_{\lam_1\lam_2}\:=
\frac{\sqrt{2}}{C}
\left(
\begin{array}{cc}
m +\frac{{\bf k}^2_\perp}{M_0+ 2m} &
k^R\frac{x_1M_{0} + m}{M_0 + 2m}\\
-k^R\frac{x_2M_0 + m}{M_0 + 2m} &
-\frac{(k^R)^2}{M_0 + 2m}
      \end{array}
    \right),
\ee
where
$C= \sqrt{2x_1x_2}M_0$.
For the radial wave function $\phi_R$, we use the same Gaussian wave function
for both pseudoscalar and vector mesons
\be\label{rad}
\phi_R(x_i,{\bf k}_{i\perp})=\frac{4\pi^{3/4}}{\beta^{3/2}}
\sqrt{\frac{\partial k_z}{\partial x}}
{\rm exp}(-{\vec k}^2/2\beta^2),
\ee
where $\beta$ is the variational parameter.
When the longitudinal component $k_z$ is defined by
$k_z=(x-1/2)M_0$, the Jacobian of the variable
transformation $\{x,{\bf k}_\perp\}\to k=({\bf k}_\perp, k_z)$
is given by $\partial k_z/\partial x=M_0/(4x_1x_2)$.
Also,
the normalization factor in Eq.~(\ref{rad}) is obtained from 
the total wave function normalization given by
\be\label{norm}
\int^1_0dx\int\frac{d^2{\bf k}_\perp}{16\pi^3}
|\Psi^{JJ_z}_{100}(x,{\bf k}_{\perp},\lam_1\lam_2)|^2=1.
\ee
The quark distribution amplitude(DA) of a hadron 
in our LFQM can be obtained from the hadronic
wave function by integrating out the transverse momenta of the quarks
in the hadron,
\bea\label{DA}
\phi(x,\mu)&=&\int^{{\bf k}^2_\perp<\mu^2}
\frac{d^2{\bf k}_\perp}{16\pi^3}
\Psi^{JJ_z}_{100}(x,{\bf k}_\perp,\lam_1\lam_2),
\eea
where $\mu$ denotes the separation scale between the perturbative and 
nonperturbative regimes.
The dependence on the scale $\mu$ is then given by the QCD
evolution equation~\cite{BL} and can be calculated
perturbatively. However, the distribution amplitudes at a certain low
scale can be obtained by the necessary nonperturbative input from LFQM.
The presence of the damping Gaussian factor in our LFQM allows
us to perform the integral up
to infinity without loss of accuracy. The quark DAs for $\eta_c$ 
and $J/\psi$ mesons are constrained by
\bea\label{DA_norm}
\int^1_0\phi_{\eta_c(J/\psi)}(x,\mu)dx
=\frac{f_{\eta_c(J/\psi)}}{2\sqrt{6}},
\eea
where the decay constant is defined as
\bea\label{fp}
\la 0|\bar{q}\gamma^\mu\gamma_5 q|\eta_c\ra=if_{\eta_c} P^\mu,
\eea
for a $\eta_c$ meson and
\bea\label{fv}
\la 0|\bar{q}\gamma^\mu q|J/\psi(P,h)\ra&=&f_{J/\psi} M_{J/\psi}\ep^\mu(h),
\nonumber\\
\la 0|\bar{q}\sigma^{\mu\nu} q|J/\psi(P,h)\ra
&=&if^T_{J/\psi} [\ep^\mu(h) P_\nu -\ep^\nu(h) P_\mu],
\nonumber\\
\eea
for a $J/\psi$ meson with longitudinal($h=0$) and
transverse($h=\pm 1$) polarizations, respectively.
The constraint of Eq.~(\ref{DA_norm}) must be
independent of cut-off $\mu$ up to corrections of order $\Lambda^2/\mu^2$,
where $\Lambda$ is some typical hadronic scale($< 1$ GeV)~\cite{BL}.
For the nonperturbative valence wave
function given by Eq.~(\ref{rad}), we take
$\mu\sim m_c$ as an optimal scale for our LFQM description of $J/\psi$ and $\eta_c$.

The explicit form of the $\eta_c$ decay constant is given by~\cite{CJ_DA}
\bea\label{fp_LFQM}
\frac{f_{\eta_c}}{2\sqrt{6}}
=\int^1_0 dx\int \frac{d^2{\bf k}_\perp}{16\pi^3}
\frac{m}{\sqrt{m^2 + {\bf k}^2_\perp}}\phi_R(x,{\bf k}_\perp).
\eea
The decay constants for the longitudinally
and transversely polarized $J/\psi$ meson are given by~\cite{CJ_DA}
\begin{equation}\label{fv_LFQM}
\frac{f_{J/\psi}}{2\sqrt{6}}
=\int^1_0 dx\int \frac{d^2{\bf k}_\perp}{16\pi^3}
\frac{\phi_R(x,{\bf k}_\perp)}{\sqrt{m^2 + {\bf k}^2_\perp}}
\biggl[ m + \frac{2{\bf k}^2_\perp}{M_0+2m}\biggr],
\end{equation}
\be\label{fvT_LFQM}
\frac{f^T_{J/\psi}}{2\sqrt{6}}
=\int^1_0 dx\int\frac{d^2{\bf k}_\perp}{16\pi^3}
\frac{\phi_R(x,{\bf k}_\perp)}{\sqrt{m^2 + {\bf k}^2_\perp}}
\biggl[ m + \frac{{\bf k}^2_\perp}{M_0+2m}\biggr],
\ee
respectively. While the constant $f_{J/\psi}$ is known
from the experiment, the constant $f^T_{J/\psi}$ is not that easily accessible in
experiment but can be estimated theoretically.

We may also redefine the quark DA as
$\Phi_{\eta_c(J/\psi)}(x)=(2\sqrt{6}/f_{\eta_c(J/\psi)})\phi(x)$ for the
normalization given by
\bea\label{DA2}
\int^1_0\Phi_{\eta_c(J/\psi)}(x)dx=1.
\eea
The quark DA $\Phi(x)$ evolved in the leading order of $\alpha_s(\mu)$
is usually expanded in
Gegenbauer polynomials $C^{3/2}_n$ as
\bea\label{Gegen}
\Phi(x,\mu)&=&\Phi_{\rm as}(x)\biggl[
1+\sum_{n=1}^{\infty}a_n(\mu)C^{3/2}_n(2x-1) \biggr],
\eea
where $\Phi_{\rm as}(x)=6x(1-x)$ is the asymptotic DA and the coefficients
$a_n(\mu)$ are Gegenbauer moments~\cite{BL}.
The Gegenbauer moments with $n>0$ describe
how much the DAs deviate from the asymptotic one. 
In addition to the Gegenbauer moments,
we can also define
the expectation value of the longitudinal momentum, so-called
$\xi$-moments:
\bea\label{xi_mom}
\la\xi^n\ra&=&\int^1_{-1}d\xi \xi^n\hat{\Phi}(\xi)
=\int^{1}_{0}dx \xi^n\Phi(x),
\eea
where $\Phi(x)=2\hat{\Phi}(2x-1)$ normalized by $\la\xi^0\ra=1$.

The $\xi$ moments are related to the Gegenbauer moments 
as follows (up to $n=6$):
\bea\label{xia}
\la\xi^2\ra &=& \frac{1}{5} + a_2\frac{12}{25},
\nonumber\\
\la\xi^4\ra&=&\frac{3}{35} + a_2\frac{8}{35} + a_4\frac{8}{77},
\nonumber\\
\la\xi^6\ra&=&\frac{1}{21} + a_2\frac{12}{77} + a_4\frac{120}{1001}
+ a_6\frac{64}{2145}.
\eea
\section{Hard contributions to $e^+e^-\to J/\psi+\eta_c$ process}
For the exclusive process
\bea\label{CS}
e^+e^-\to \gamma^*(q)\to J/\psi(P_V) + \eta_c(P_P),
\eea
the form factor is defined as
\be\label{ff}
\la J/\psi(P_V,h)\eta_c(P_P)|J^\mu_{\rm em}|0\ra
=\epsilon^{\mu\nu\rho\sigma}\epsilon^*_\nu P_{V\rho}P_{P\sigma}
{\cal F}(q^2),
\ee
where $\epsilon^*_\nu(P_V,h)$ is the polarization vector of the vector
meson with four momentum $P_V$ and helicity $h$.
The cross section can be calculated as
\be\label{cross}
\sigma(e^+e^-\to J/\psi\eta_c)
= \frac{\pi\alpha^2}{6}|{\cal F}(s)|^2
\biggl(1 - \frac{4M^2_h}{s}\biggr)^{3/2},
\ee
where we neglect the small mass difference between $J/\psi$ and $\eta_c$, i.e.
$M_h\approx M_{J/\psi}\approx M_{\eta_c}$.

\begin{figure}
\includegraphics[width=2.0in,height=2.0in]{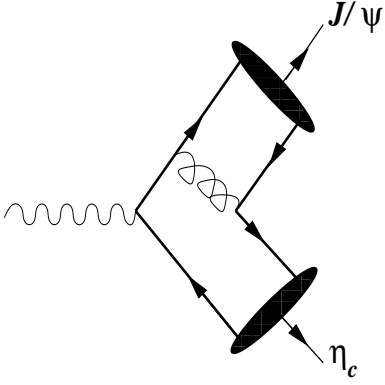}
\caption{One of the four Feynman diagrams for the amplitude.}
\label{fig1}
\end{figure}
At leading order of $\al_s$, the contribution to the form factor comes from
four Feynman diagrams; one of them is shown in Fig.~\ref{fig1}.
To obtain the timelike form factor ${\cal F}(q^2)$ 
for the process $e^+e^-\to\gamma^*\to J/\psi\eta_c$, we first
calculate the radiative decay process $\eta_c(P)+\gamma^*(q)\to J/\psi(P')$ 
using the Drell-Yan-West($q^+=q^0+q^3=0$) frame 
\bea\label{mom}
P &=&\biggl(P^+, \frac{M^2_{\eta_c}}{P^+},{\bf 0}_\perp\biggr),
P' = \biggl(P^+, \frac{M^2_{J/\psi}+{\bf q}^2_\perp}{P^+}, 
{\bf q}_\perp\biggr),
\nonumber\\
q&=&\biggl(0, \frac{{\bf q}^2_\perp}{P^+},{\bf q}_\perp\biggr),
\eea
where the four momentum transfer is spacelike, i.e.
$q^2=q^+q^- - {\bf q}^2_\perp=-{\bf q}^2_\perp<0$. We then analytically
continue the spacelike form factor ${\cal F}({\bf q}^2_\perp)$ 
to the timelike $q^2>0$ region by changing $-{\bf q}^2_\perp$ to
$q^2$ in the form factor.

In the calculations of the form factor ${\cal F}(q^2)$, we use the
`+'-component of currents and the transverse($h=\pm 1$) polarization
for $J/\psi$
given by
\bea\label{pol}
\ep^*(\pm1)&=&\mp\frac{1}{\sqrt{2}}\biggl(0,\frac{2q^L}{P^+_1},1,
\mp i\biggr),
\eea
where $q^L=q_x -iq_y$.
For the longitudinal($h=0$) polarization, it is hard to extract the form
factor since both sides of Eq.~(\ref{ff}) vanish for any $q^2$ value.
In the energy region where PQCD is applicable, the hadronic matrix element 
$\la J/\psi|J^+_{\rm em}|\eta_c\ra$ can be calculated within the leading 
order PQCD by means of a homogeneous Bethe-Salpeter(BS) equation for the 
meson wave function. 
Formally, one may consider a contribution even at lower order without 
any gluon exchange as often called Feynman mechanism. However, we don't 
need to take this contribution into account in this work because we are 
considering the production process of heavy mesons that ought to require 
high momentum transfer between the primary quark-antiquark pair production 
and the secondary quark-antiquark pair production in order to get the 
final state heavy mesons. Since the final bound state wavefunctions satisfy 
the BS type iterative bound state equation, 
one gluon exchange can be generated by iteration from the wavefunction 
part even if the scattering amplitude formally has no gluon exchange.
We thus generate the hard gluon exchange from the iteration of bound 
state wavefunction and consider the leading order PQCD contribution 
in the framework of LFD. The secondary quark-antiquark 
pair production can occur only through the gluon momentum transfer 
due to the rational light-front energy-momentum dispersion relation.
The quark-antiquark pair production from the vacuum is suppressed 
in the LFD and the absence of zero-mode contribution 
can be shown by the direct power counting method that we presented in 
our previous work of weak transition form factors between pseudoscalar 
and vector mesons~\cite{CJadd}.
Taking the perturbative kernel of the BS equation as a
part of hard scattering amplitude $T_H$, one thus obtains 
\begin{eqnarray}\label{F_PQCD}
\la J/\psi|J^+_{\rm em}|\eta_c\ra
&=&\sum_{\lam,\lam'}
\int [d^3k][d^3l] \Psi^{11\dagger}_{100}(y,{\bf l}_\perp,\lam)
\nonumber\\
&&\times
T_H(x, {\bf k}_\perp; y,{\bf l}_\perp;{\bf q}_\perp;\lam,\lam')
\nonumber\\
&&\times \Psi^{00}_{100}(x,{\bf k}_\perp,\lam'),
\end{eqnarray}
where $[d^3k]=dxd^2{{\bf k}_\perp}/16\pi^3$ and
$T_H$ contains all two-particle irreducible amplitudes
for $\gamma^* + q{\bar q}\to q{\bar q}$ from the iteration
of the LFQM wave function with the BS kernel.
On the other hand, the right-hand-side of Eq.~(\ref{ff}) for the matrix element
of $J^+$ is obtained as
\bea\label{Jplus}
\la J/\psi|J^+_{\rm em}|\eta_c\ra
&=&\frac{P^+}{\sqrt{2}}q^L {\cal F}(q^2).
\eea
Here, we set $P^+=1$ without any loss of generality.
Therefore, we get the form factor as
\bea\label{calF}
{\cal F}(q^2) &=&
\int [d^3k][d^3l]\phi_R(y,{\bf l}_\perp) {\cal T}_H
\phi_R(x,{\bf k}_\perp),
\eea
where we combined the spin-orbit
wave function into the original $T_H$ to form a new ${\cal T}_H$, i.e.
\bea\label{newT}
{\cal T}_{H}&=&\frac{\sqrt{2}}{q^L}
\sum_{\lam,\lam'}
{\cal R}^{11\dagger}_{\lam'_1\lam'_2}(y,{\bf l}_\perp)
T_H(x, {\bf k}_\perp; y,{\bf l}_\perp;{\bf q}_\perp;\lam,\lam')
\nonumber\\
&&\times {\cal R}^{00}_{\lam_1\lam_2}(x, {\bf k}_\perp).
\eea
We should note that the form factor ${\cal F}(q^2)$ has a dimension
[1/GeV] so that the cross section has a dimension of [barn] where
1 GeV$^{-2}$=0.39 mb in the natural unit($\hbar=c=1$). 
Since the measure $[d^3k]$ has the dimension of [GeV$^2$] and our radial 
wave function $\phi_R$ has the dimension of [1/GeV], the 
amplitude ${\cal T}_H$ has the dimension of [1/GeV$^{3}$].

\begin{figure*}
\includegraphics[width=5.0in,height=2.0in]{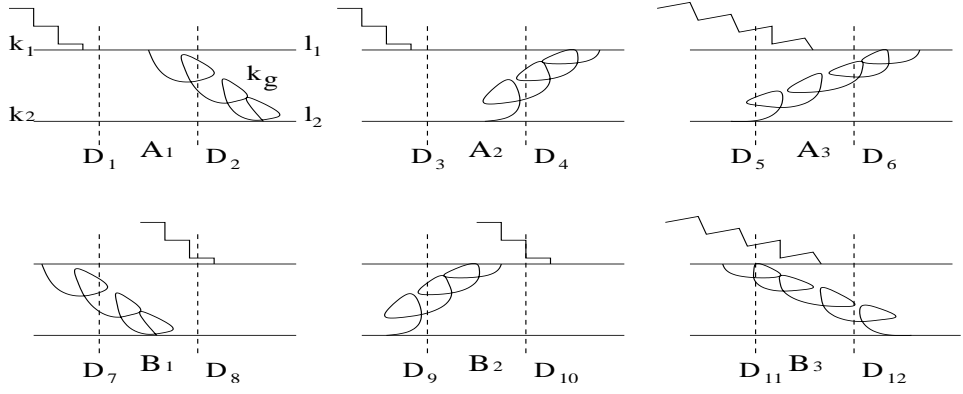}
\caption{Leading order(in $\al_s$) light-front time-ordered diagrams of 
the hard scattering amplitude for $\eta_c(P)\to\gamma^*(q)+J/\psi(P')$
process.}
\label{fig2}
\end{figure*}
The leading order light-front time-ordered diagrams for the meson
form factor are shown in Fig.~\ref{fig2}, where the energy denominators
are given by
\bea\label{EDA}
D_1&=& M^2 + {\bf q}^2_\perp
- \frac{({\bf k}_\perp + {\bf q}_\perp)^2+m^2}{x_1}
-\frac{{\bf k}^2_\perp + m^2}{x_2},
\nonumber\\
D_2&=& M^2+ {\bf q}^2_\perp
- \frac{(y_1{\bf q}_\perp + {\bf l}_\perp)^2+m^2}{y_1}
-\frac{{\bf k}^2_\perp + m^2}{x_2}
\nonumber\\
&&-\frac{(y_2{\bf q}_\perp + {\bf k}_\perp - {\bf l}_\perp)^2}{y_2-x_2},
\nonumber\\
D_3 &=& D_1,
\nonumber\\
D_4&=& M^2+ {\bf q}^2_\perp
- \frac{({\bf k}_\perp + {\bf q}_\perp)^2+m^2}{x_1}
\nonumber\\
&&-\frac{(y_2{\bf q}_\perp + {\bf k}_\perp - {\bf l}_\perp)^2}{x_2-y_2}
-\frac{(y_2{\bf q}_\perp - {\bf l}_\perp)^2 + m^2}{y_2},
\nonumber\\
D_5 &=& M^2 - \frac{{\bf k}_\perp^2 + m^2}{x_1}
- \frac{(y_2{\bf q}_\perp + {\bf k}_\perp - {\bf l}_\perp)^2}{x_2-y_2}
\nonumber\\
&&- \frac{(y_2{\bf q}_\perp - {\bf l}_\perp)^2 + m^2}{y_2},
\nonumber\\
D_6 &=& D_4.
\nonumber\\
D_7 &=& M^2
- \frac{({\bf l}_\perp - y_2{\bf q}_\perp)^2+m^2}{y_1}
-\frac{{\bf k}^2_\perp + m^2}{x_2}
\nonumber\\
&&-\frac{(y_2{\bf q}_\perp + {\bf k}_\perp - {\bf l}_\perp)^2}{y_2-x_2},
\nonumber\\
D_8 &=& M^2
- \frac{({\bf l}_\perp - y_2{\bf q}_\perp)^2+m^2}{y_1}
- \frac{({\bf l}_\perp - y_2{\bf q}_\perp)^2+m^2}{y_2},
\nonumber\\
D_9 &=& D_5,
D_{10}= D_8,
D_{11}= D_7,
D_{12}= D_2.
\eea
According to the rules of light-front perturbation theory, the
hard scattering amplitude for the diagram $A_1$ in Fig.~\ref{fig2}
is given by
\bea\label{TA1}
T_{A_1}&=& (-1)\frac{\theta(k^+_g)}{k^+_g D_1D_2}
\bar{u}(k_1 + q)\gamma^+ u(k_1)
\nonumber\\
&&\times\biggl[4\pi\al_s C_F\bar{u}(l_1)\gamma^\mu u(k_1+q)
d^{(k_g)}_{\mu\nu}\bar{v}(k_2)\gamma^\nu v(l_2)\biggr]
\nonumber\\
&=& (-2)(4\pi\al_s C_F)\frac{\theta(k^+_g)}{k^+_g D_1D_2}N_{A_1}
\eea
where $N_{A_1}= \bar{u}(l_1)\gamma^\mu u(k_1+q)
d^{(k_g)}_{\mu\nu}\bar{v}(k_2)\gamma^\nu v(l_2)$ is the gluon-fermion 
vertex part with the gauge dependent polarization sum $d_{\mu\nu}$ for 
the gluon, $C_F=4/3$ is the color factor
and the notation $u(p)$ denotes actually $u(p)/\sqrt{p^+}$ for the internal 
fermions in a scattering amplitude. In the Feynman gauge, the polarization 
sum $d^{\mu\nu}$ equals to $g^{\mu\nu}$. In the light-front gauge 
$\eta\cdot A=A^+=0$,
\bea\label{dmn}
d^{(k_g)}_{\mu\nu}&=& 
\sum_{\lambda=1,2}\epsilon_\mu(k_g,\lambda) \epsilon_\nu(k_g,\lambda)
\nonumber\\
&=& -g_{\mu\nu} + 
\frac{\eta_\mu (k_g)_\nu + \eta_\nu (k_g)_\mu}{k^+_g},
\eea
where $k_g\cdot\epsilon=\eta\cdot\epsilon=0$.
Since the gluon propagator has an
instantaneous part($\eta^\mu\eta^\nu/(k^+_g)^2$ in the light-front gauge), 
we absorb this instantaneous
contribution into the regular propagator by replacing $k_g$ by  
$ {\bar k}_g=(y_2-x_2,{\bar k}^-_g, 
y_2{\bf q}_\perp + {\bf k}_\perp -{\bf l}_\perp)$,
where ${\bar k}_g^-=P^-+q^--l^-_1-k^-_2$ includes the instantaneous 
contribution.\\
\noindent
Similarly, the hard scattering amplitude for the diagram 
$A_2$ in Fig.~\ref{fig2} is given by
\bea\label{TA2}
T_{A_2}&=& (-2)(4\pi\al_s C_F)\frac{\theta(k^+_g)}{k^+_g D_3D_4}
N_{A_2},
\eea
where $N_{A_2}$ has the same form as $N_{A_1}$ but with different
gluon momentum 
${\bar k}_g=(x_2-y_2,{\bar k}_g^-, {\bf l}_\perp - y_2 {\bf q}_\perp 
- {\bf k}_\perp)$. The diagram $A_2$ has also the instantaneous contribution 
and absorb it into ${\bar k}_g^-=P^-+q^- -(k_1 + q)^- - l^-_2$.\\
\noindent
The hard scattering amplitude for the diagram $A_3$ in Fig.~\ref{fig2} is 
given by
\bea\label{TA3}
T_{A_3}&=& (-2)(4\pi\al_s C_F)\frac{\theta(k^+_g)}{k^+_g D_5D_6}
N_{A_3},
\eea
where $N_{A_3}$ has also the same form as $N_{A_1}$ or $N_{A_2}$. However, 
in this case we have just regular gluon
propagator with the four momentum
$k_g=(x_2-y_2,k_g^-, {\bf l}_\perp - y_2 {\bf q}_\perp - {\bf k}_\perp)$
and $k_g^-={\bf k}^2_{g\perp}/k^+_g$ since the
diagram does not have the instantaneous part.\\
\noindent
Likewise, one can obtain the hard scattering amplitudes corresponding to 
the diagrams $B_i$.

If one includes the higher twist effects such as intrinsic transverse momenta
and the quark masses, the LF gauge part proportional to
$ 1/ k^+_g$ leads to  a singularity although the Feynman gauge part
$g_{\mu\nu}$ gives the regular amplitude. This is due to the gauge-invariant
structure of the amplitudes.
The covariant derivative $D_\mu = \partial_\mu + igA_\mu$ makes both the
intrinsic transverse momenta, ${\bf k}_\perp$ and
${\bf l}_\perp$, and the transverse gauge degree of freedom $g{\bf A}_\perp$
be of the same order, indicating the need of the higher Fock state 
contributions to ensure the gauge invariance~\cite{JPS}.  
However, we can show that
the sum of six diagrams for the LF gauge part($1/k^+_g$ terms)
vanishes in the limit that the LF energy differences
$\Delta_x$ and $\Delta_y$ go to zero, where $\Delta_x$ and
$\Delta_y$ are given by
\bea\label{zb}
\Delta_x &=& M^2 - \frac{{\bf k}^2_\perp + m^2}{x_1x_2} =M^2 - M^2_{0x},
\nonumber\\
\Delta_y &=& M^2 - \frac{{\bf l}^2_\perp + m^2}{y_1y_2} =M^2-M^2_{0y}.
\eea
Details of the proof can be found in our previous work~\cite{CJD}.
In this work, we follow the same procedure presented in Ref.~\cite{CJD}
and calculate the higher twist effects
in the limit of $\Delta_x = \Delta_y = 0$ to avoid
the involvement of the higher Fock state contributions. Our limit
$\Delta_x=\Delta_y=0$(but $\sqrt{\la{\bf k}^2_\perp\ra}=\beta\neq 0$)
may be considered as a zeroth order
approximation in the expansion of a scattering amplitude. That is,
the scattering amplitude $T_H$ may
be expanded in terms of LF energy difference $\Delta$ as
$T_H = [T_H]^{(0)}
+ \Delta[T_H]^{(1)} + \Delta^2[T_H]^{(2)}+\cdots$,
where $[T_H]^{(0)}$ corresponds to
the amplitude in the zeroth order of $\Delta$.
This approximation should be distinguished from the
zero-binding(or peaking) approximation that corresponds
to $M=m_1 + m_2$ and ${\bf k}_\perp=\beta=0$.
The point of this distinction is to note that $[T_H]^{(0)}$ includes the
binding energy effect(i.e. ${\bf k}_\perp,{\bf l}_\perp\neq 0$)
that was neglected in the peaking approximation.

In zeroth order of $\Delta$, one can show that the energy denominators 
entering in the diagrams $A_i$ and $B_i(i=1,2,3)$ have the following relations
\bea\label{Di}
{\cal D}_2 + {\cal D}_4 &=& {\cal D}_1,
{\cal D}_2 + {\cal D}_5 = 0,
{\cal D}_2 + {\cal D}_8 = {\cal D}_7,
\eea
where ${\cal D}_i={\rm lim}_{\Delta_x=\Delta_y=0}D_i$ and  
\bea\label{D128}
{\cal D}_1&=& -(x_2{\bf q}^2_\perp + 2{\bf k}_\perp\cdot {\bf q}_\perp)/ x_1,
\nonumber\\
{\cal D}_8&=&-(y_2{\bf q}^2_\perp - 2{\bf l}_\perp\cdot {\bf q}_\perp)/ y_1,
\nonumber\\
{\cal D}_2&=& -[x^2_2y^2_2{\bf q}^2_\perp
+ x^2_2{\bf l}^2_\perp + y^2_2 {\bf k}^2_\perp
-2x_2y_2(x_2 {\bf l}_\perp\cdot {\bf q}_\perp
\nonumber\\
&&- y_2 {\bf k}_\perp\cdot {\bf q}_\perp
+ {\bf k}_\perp\cdot {\bf l}_\perp) + (y_2-x_2)^2m^2]
\nonumber\\
&&\;\;/[x_2y_2(y_2-x_2)].
\eea

As one can see from Eqs.~(\ref{R00}) and (\ref{R11}), 
the leading twist(LT)(i.e. neglecting tranverse momenta ${\bf k}_\perp$ and 
${\bf l}_\perp$) helicity contributions for 
$\eta_c(P)+\gamma^*(q)\to J/\psi(P',\ep(+1))$ process
come from two leading helicity 
$\Delta H=|\lam_{J/\psi}-\lam_{\eta_c}|=1$ components, i.e.
$\up\down\to\up\up$ and $\down\up\to\up\up$ as in the nonrelativistic
spin case, i.e. $\frac{1}{\sqrt{2}}(|\up\down\ra - |\down\up\ra)$
for $\eta_c$ meson to $|\up\up\ra$ for $J/\psi$ meson.

%
%
\begin{table*}
\caption{Leading helicity contributions to the hard scattering
amplitudes, $T^{(\up\down\to\up\up)}_H$ and $T^{(\down\up\to\up\up)}_H$.
The momentum variable $p^{R(L)}$ represents $p^{R(L)}=p_x\pm ip_y$.}
\label{t1}
\begin{tabular}{ccccc} \hline\hline
$|\Delta H=1|$ & $N_{A_1}=N_{A_2}=N_{A_3}$ &
$\sum^3_{i=1}T_{A_i}(\Delta_x=\Delta_y=0)$ &
$N_{B_1}=N_{B_2}=N_{B_3}$ &
$\sum^3_{i=1}T_{B_i}(\Delta_x=\Delta_y=0)$ \\
\hline
$\up\down\to\up\up$
&  $\frac{2m(x_1 l - y_1 k - x_2y_1 q)^L}{x_2y_1y_2}$
& $-\frac{8\pi\al_s C_F}{(y_2-x_2)
{\cal D}_1{\cal D}_2}N_{A_1}^{(\up\down\to\up\up)}$ &
$\frac{2m(x_1 l - y_1 k - x_1y_2 q)^L}{x_2y_1y_2}$
& $-\frac{8\pi\al_s C_F}{(y_2-x_2)
{\cal D}_2{\cal D}_8}N_{B_1}^{(\up\down\to\up\up)}$
\\
$\down\up\to\up\up$
& $-\frac{2m(x_2 l - y_2 k - x_2y_2 q)^L}{x_1y_1y_2}$
& $-\frac{8\pi\al_s C_F}{(y_2-x_2)
{\cal D}_1{\cal D}_2}N_{A_1}^{(\down\up\to\up\up)}$&
$-\frac{2m(x_2 l - y_2 k - x_2y_2 q)^L}{x_1y_1y_2}$
& $-\frac{8\pi\al_s C_F}{(y_2-x_2)
{\cal D}_2{\cal D}_8}N_{B_1}^{(\down\up\to\up\up)}$\\
\hline\hline
\end{tabular}
\end{table*}
In Table~\ref{t1}, we summarize our results for the hard scattering amplitude
$T_H$ for these two leading helicity $\Delta H=1$ components
in zeroth order of binding energy limit.
For instance, $\sum^3_{i=1}T^{(\up\down\to\up\up)}_{A_i}$ and
$\sum^3_{i=1}T^{(\up\down\to\up\up)}_{B_i}$ are obtained as
\begin{widetext}
\bea\label{th_A}
\biggl[\sum_iT_{A_i}^{(\up\down\to\up\up)}\biggr]_{\Delta=0}
&=&-8\pi\al_s C_FN^{(\up\down\to\up\up)}_{A_1}\biggl[
\frac{\theta(y_2-x_2)}{(y_2-x_2){\cal D}_1{\cal D}_2}
+\frac{\theta(x_2-y_2)}{(x_2-y_2)}
\biggl(\frac{1}{{\cal D}_3{\cal D}_4}+\frac{1}{{\cal D}_5{\cal D}_6}\biggr)
\biggr]
\nonumber\\
&=&-8\pi\al_s C_FN^{(\up\down\to\up\up)}_{A_1}
\biggl[\frac{1}{(y_2-x_2){\cal D}_1{\cal D}_2}\biggr],
\nonumber\\
\biggl[\sum_iT_{B_i}^{(\up\down\to\up\up)}\biggr]_{\Delta=0}
&=&-8\pi\al_s C_FN^{(\up\down\to\up\up)}_{B_1}\biggl[
\frac{\theta(y_2-x_2)}{(y_2-x_2)}
\biggl(\frac{1}{{\cal D}_7{\cal D}_8}+\frac{1}{{\cal D}_{11}{\cal D}_{12}}
\biggr)
+\frac{\theta(x_2-y_2)}{(x_2-y_2){\cal D}_9{\cal D}_{10}}
\biggr]
\nonumber\\
&=&-8\pi\al_s C_FN^{(\up\down\to\up\up)}_{B_1}
\biggl[\frac{1}{(y_2-x_2){\cal D}_2{\cal D}_8}\biggr].
\eea
\end{widetext}
To derive the final results in Eq.~(\ref{th_A}),
we use the following identities obtained from  Eqs.~(\ref{EDA}) and
(\ref{Di}):
\bea\label{DsumA}
\frac{1}{{\cal D}_3{\cal D}_4} + \frac{1}{{\cal D}_5{\cal D}_6}
&=& \frac{1}{{\cal D}_1{\cal D}_4} + \frac{1}{{\cal D}_5{\cal D}_4}
\nonumber\\
&=&\frac{ {\cal D}_1+{\cal D}_5}{{\cal D}_1{\cal D}_4{\cal D}_5}
=\frac{ {\cal D}_1 -{\cal D}_2}{{\cal D}_1{\cal D}_4(-{\cal D}_2)}
\nonumber\\
&=&\frac{-1}{{\cal D}_1{\cal D}_2},
\eea
for the diagram $A$ and
\bea\label{DsumB}
\frac{1}{{\cal D}_7{\cal D}_8} + \frac{1}{{\cal D}_{11}{\cal D}_{12}}
&=& \frac{1}{{\cal D}_7{\cal D}_8} + \frac{1}{{\cal D}_7{\cal D}_2}
\nonumber\\
&=&\frac{ {\cal D}_2+{\cal D}_8}{{\cal D}_7{\cal D}_2{\cal D}_8}
=\frac{1}{{\cal D}_2{\cal D}_8},
\nonumber\\
\frac{1}{{\cal D}_9{\cal D}_{10}}&=& \frac{1}{{\cal D}_5{\cal D}_8}
=\frac{-1}{{\cal D}_2{\cal D}_8},
\eea
for the diagram $B$. The above identities lead to 
$\theta(y_2-x_2)+\theta(x_2-y_2)=1$ in Eq.~(\ref{th_A}).

By adding all six LF time-ordered diagrams, we obtain
\be\label{th_AB}
[T_{H}^{(\up\down\to\up\up)}]^{(0)}
=
\sum^3_{i=1}\biggl[T_{A_i}^{(\up\down\to\up\up)}
+ T_{B_i}^{(\up\down\to\up\up)}\biggr]_{\Delta=0}.
\ee
Similarly, one can easily obtain the helicity $\down\up\to\up\up$ contribution
to the hard scattering amplitude, $T^{(\down\up\to\up\up)}_H$ from 
Table~\ref{t1}.

In the numerical calculations for the higher twist contributions,
one may keep effectively only the leading order of higher twist terms
such as ${\bf k}^2_\perp/{\bf q}^2_\perp$, ${\bf l}^2_\perp/{\bf q}^2_\perp$, 
and ${\bf k}_\perp\cdot{\bf l}_\perp/{\bf q}^2_\perp$
due to the fact that ${\bf k}^2_\perp\ll{\bf q}^2_\perp$ and
${\bf l}^2_\perp\ll{\bf q}^2_\perp$ in large momentum transfer region where
PQCD is applicable~\cite{CJD,HWW}. 
As shown in our previous work~\cite{CJD}, this can be done by
neglecting the subleading higher twist terms accordingly
both in the energy denominators and the numerators for the hard scattering
amplitude $T_H$.
This procedure is very similar to the recent investigation of 
the relativistic and bound state effects not based on the 
LFD but including the relativistic effects up to the second order of the
relative quark velocity, i.e. $\la v^2\ra$\cite{EM}.
Indeed, our numerical result neglecting the higher orders of 
${\bf k}^2_\perp/{\bf q}^2_\perp$, ${\bf l}^2_\perp/{\bf q}^2_\perp$ 
and ${\bf k}_\perp\cdot{\bf l}_\perp/{\bf q}^2_\perp$
is very close to that presented in Ref.\cite{EM} (see Section IV).
However, in this work, we include all higher orders of
${\bf k}^2_\perp/{\bf q}^2_\perp$, ${\bf l}^2_\perp/{\bf q}^2_\perp$
and ${\bf k}_\perp\cdot{\bf l}_\perp/{\bf q}^2_\perp$. This corresponds
to keep effectively all higher orders of the relative quark velocity
beyond $\la v^2\ra$. We compare our full result with the one 
neglecting the corrections of order ${\cal{O}}(\la v^4\ra)$.
 
Using Eq.~(\ref{newT}), we then obtain the leading helicity contributions 
to the hard scattering amplitude combined with the spin-orbit wave function
in zeroth order of $\Delta$ as follows 
\bea\label{RH_LO}
[{\cal T}_{H}]^{(0)}&=&\frac{\sqrt{2}}{q^L}
{\cal R}^{11\dagger}_{\up\up}
\biggl[T^{(\up\down\to\up\up)}_H{\cal R}^{00}_{\up\down}+
       T^{(\down\up\to\up\up)}_H {\cal R}^{00}_{\down\up}
\biggr]
\nonumber\\
&=&\frac{m}{\sqrt{m^2+{\bf k}^2_\perp}}
\frac{
\biggl[ m + \frac{{\bf l}^2_\perp}{M_{0y}+ 2m}
\biggr]}{\sqrt{m^2+{\bf l}^2_\perp}}
{\cal M}_H,
\eea
where 
\bea\label{NH}
{\cal M}_{H}&=&
\frac{16\pi\al_s C_F m}{x_1x_2y_1y_2(y_2-x_2)q^2}
\biggl[ (x^2_1+x^2_2)l^Lq^R 
\nonumber\\
&&- (x_1y_1+x_2y_2)k^Lq^R
\biggr]\biggl(\frac{1}{{\cal D}_1{\cal D}_2} 
+\frac{1}{{\cal D}_2{\cal D}_8}\biggr)
\nonumber\\
&&+\frac{16\pi\al_s C_F m}{x_1x_2y_1y_2(y_2-x_2)}
\biggl[
\frac{x_2(x_1y_1+x_2y_2)}{{\cal D}_1{\cal D}_2}
\nonumber\\
&&+\frac{y_2(x^2_1+x^2_2)}{{\cal D}_2{\cal D}_8}
\biggr],
\eea
and $l^Lq^R = {\bf l}_\perp\cdot{\bf q}_\perp 
+ i|{\bf l}_\perp\times{\bf q}_\perp|$.
Accordingly, the leading helicity contributions to the form factor lead 
to the following non-factorized form 
\bea\label{FNR}
{\cal F}(q^2) &=&
Q_c\int [d^3k][d^3l]\phi_R(x,{\bf k}_\perp) [{\cal T}_H]^{(0)}
\phi_R(y,{\bf l}_\perp)
\nonumber\\
&&+ Q_c(m_1\leftrightarrow m_2),
\nonumber\\
&=&2Q_c\int [d^3k][d^3l]\frac{m}{\sqrt{m^2+{\bf k}^2_\perp}}
\phi_R(x,{\bf k}_\perp) 
{\cal M}_H
\nonumber\\
&&\times
\frac{1}{\sqrt{m^2+{\bf l}^2_\perp}}
\biggl[m+\frac{{\bf l}^2_\perp}{M_{0y}+2m}\biggr]
\phi_R(y,{\bf l}_\perp),
\nonumber\\
\eea
where $Q_c=2/3$ is the charge fraction of charm quark in the unit of $e$.

In the leading twist(LT) limit neglecting the transverse momenta, 
the hard scattering amplitude in Eq.~(\ref{NH}) is reduced to 
\be\label{TH_LT}
{\cal M}^{\rm LT}_H
=\frac{16\pi\al_s C_F m}{x_1x^2y_1y^2_2q^4}
[ x_1(x_1y_1+x_2y_2)+y_1(x^2_1+x^2_2)].
\ee
Then, the form factor in Eq.~(\ref{FNR}) 
factorizes into the convolution of the nonperturbative
valence quark DAs $\phi_{\eta_c(J/\psi)}(x,\mu)$ with the perturbative 
hard scattering amplitude ${\cal M}^{\rm LT}_H$:
\bea\label{F_fac}
{\cal F}^{\rm LT}(q^2) &=& 2Q_c\int^1_0 dx\int^1_0 dy
\phi_{\eta_c}(x,\mu)
\nonumber\\
&&\times{\cal M}^{\rm LT}_H(x_i,y_i,q^2)
\phi^T_{J/\psi}(y,\mu),
\eea
where $\phi^T_{J/\psi}(y,\mu)$ is the quark DA for the transversely polarized
$J/\psi$ meson.
Furthermore, in nonrelativistic QCD(NRQCD) limit(i.e. peaking approximation) 
where the longitudinal momentum 
fractions are given by $x_i=y_i\approx 1/2$ with $M_h\approx 2m_c$,
the quark DAs for both $\eta_c$ and $J/\psi$ mesons become $\delta$-type
functions, i.e.
\bea\label{del_PV}
\phi^{\delta}_{\eta_c(J/\psi)}(x,\mu)&=&\frac{f_{\eta_c(J/\psi)}}{2\sqrt{6}}
\delta(x-1/2),
\eea
and the form factor in NRQCD limit is reduced to 
\bea\label{T_peak}
{\cal F}^{\delta}(q^2)
&\approx& 2Q_c\frac{f_{\eta_c}f_{J/\psi}}{(2\sqrt{6})^2}
\frac{256\pi\al_s(\mu) C_F}{q^4}M_h,
\eea
where the superscript $\delta$ 
for the quark DA in Eq.~(\ref{del_PV}) and the
form factor in Eq.~(\ref{T_peak}) 
represents the NRQCD result. Our NRQCD result~\footnote{
Through the private communication with Jungil Lee and Stan Brodsky, 
we have indeed confirmed the agreement between the NRQCD result
and the peaking approximation result of PQCD.} 
is exactly the same as 
that derived from Ma and Si in Ref.~\cite{Ma}(see Eqs.~(16) and (21) 
in~\cite{Ma}).

Other subleading helicity contributions to the hard scattering
amplitude that show up as next-to-leading order in transverse momenta
are summarized in Tables~\ref{t4} and~\ref{t5} of the Appendix A. 

\section{Numerical Results}
In our numerical calculations, we use our LFQM~\cite{CJ1,CJ2} parameters 
$(m_c,\beta_{cc})$ obtained from the meson spectroscopy with the variational
principle for the QCD motivated effective Hamiltonian. 
In our LFQM, we have used the two interaction potentials $V_{Q\bar{Q}}$ for $\eta_c$ 
and $J/\psi$ mesons: (1) Coulomb plus harmonic oscillator(HO) potential, and
(2) Coulomb plus linear confining potential. In addition, the hyperfine 
interaction essential for the distinction between $J/\psi$ and $\eta_c$  mesons
is included for both cases (1) and (2), viz.,
\be\label{Vqq}
V_{Q\bar{Q}}= a + V_{\rm conf} -\frac{4\al_s}{3r}
+\frac{32\pi}{9}\frac{\al_s}{m^2_c}\vec{S}_Q\cdot\vec{S}_{\bar{Q}}
\delta^3(\vec{r}),
\ee
where $V_{\rm conf}= br^2$ for the HO potential and $br$
for the linear confining potential, respectively. 
For the linear confining
potential, we use the string tension $b=0.18$ GeV$^2$, which is rather
well known from other quark-model analysis commensurate with the Regge
phenomenology~\cite{GI}. The other potential model parameters are then
fixed by the variational principle for the central Hamiltonian with
respect to the Gaussian parameter $\beta$. For instance, the model 
parameters for the linear confining potential are obtained as
$a=-0.724$ GeV, $m_c=1.8$ GeV, and the strong coupling constant 
$\al_s(\mu)=0.313$ defined by
\be\label{als}
\al_s(\mu)=\frac{12\pi}{(33-2N_f){\rm ln}(\mu^2/\Lambda^2)},
\ee
where $N_f=4$ is the number of active flavors($u,d,s$ and $c$). 
At scale $\mu\simeq m_c$ for charmonium, our value of
$\Lambda=162$ MeV, the scale asscociated with nonperturbative effects involving
light quarks and gluons, is consistent with the usual $\Lambda_{QCD}\simeq 200$ MeV. 
Our value of $\al_s=0.313$ is also quite comparable
with other quark model predictions such as 0.35, 0.45, $0.30\sim0.38$, and
0.314 from ISGW2 model~\cite{ISGW2}, Cornell potential model~\cite{Cor},
Bodwin-Kang-Lee(BKL) model~\cite{BKL}(in next-to-leading order in $\al_s$), 
and relativistic quark model~\cite{EM}, respectively. 
Lattice measurements of the heavy-quark potential yield the values for
effective coupling $\al_s$ of 0.22 in the quenched case and approximately
0.26 in the unquenched case~\cite{Bali}. The HO potential model parameters
are obtained in a similar way as in the case of the linear potential. We should note that 
the root-mean-square value of the transverse momentum in our LFQM is equal to the 
Gaussian $\beta$ value, i.e.  $\sqrt{\la{\bf k}^2_\perp\ra_{cc}}=\beta_{cc}$.

\begin{table}[t]
\caption{Decay constants[MeV] of $\eta_c$ and $J/\psi$ obtained
from our variational parameters ($m_c=1.8,\beta$)[GeV] and
compared with the experimental data.}\label{t2}
\begin{tabular}{ccccc} \hline\hline
 & Linear & HO & HO$^\prime$ & Exp. \\
 & ($\beta=0.6509$) &  ($\beta=0.6998$) &
 ($\beta=0.7278$) &  \\
\hline
$f_{\eta_c}$ & 326 & 354 & 370 & $335\pm75$~\cite{CLEO01} \\
\hline
$f_{J/\psi}$ & 360 & 395 & 416 & $416\pm 6$~\cite{PDG06}\\
\hline
$f^T_{J/\psi}$ & 343 & 375 & 393 & $-$\\
\hline\hline
\end{tabular}
\end{table}
In Table~\ref{t2}, we summarize our results for decay constants of
$\eta_c$ and $J/\psi$ obtained from our variational parameters
($m_c=1.8$ GeV, $\beta_{cc}=$0.6509 GeV) for the linear
potential[second column] and ($m_c=1.8$ GeV, $\beta_{cc}=$0.6998 GeV)
for the HO potential[third column]. Our results for the decay constants 
$f_{\eta_c}=326[354]$ MeV, $f_{J/\psi}=360[395]$ MeV, and
$f^T_{J/\psi}=343[375]$ MeV obtained from the linear[HO] potential parameters 
are quite comparable with the current experimental
data, $(f_{\eta_c})_{\rm exp}=335\pm 75$ MeV~\cite{CLEO01} and 
$(f_{J/\psi})_{\rm exp}=416\pm 6$ MeV~\cite{PDG06} 
as well as other theoretical model calculations such as the QCD sum 
rules~\cite{BLL,Bra} in which the
decay constants were obtained as $f_{\eta_c}=346$ MeV, $f_{J/\psi}=412$ MeV and
$f^T_{J/\psi}=409$ MeV.
As a sensitivity check of our variational parameters, we include 
in Table~\ref{t2} another Gaussian parameters($m_c=1.8$ GeV, 
$\beta_{cc}=$0.7278 GeV) to fit the central value of the experimental $J/\psi$
decay constant. We denote this as HO$^\prime$ in Table~\ref{t2}.
In the following numerical calulations, we present all of these three cases
(linear, HO, HO$^\prime$) to show the parameter sensitivity of our results.

\begin{figure}[t]
\includegraphics[width=3.0in]{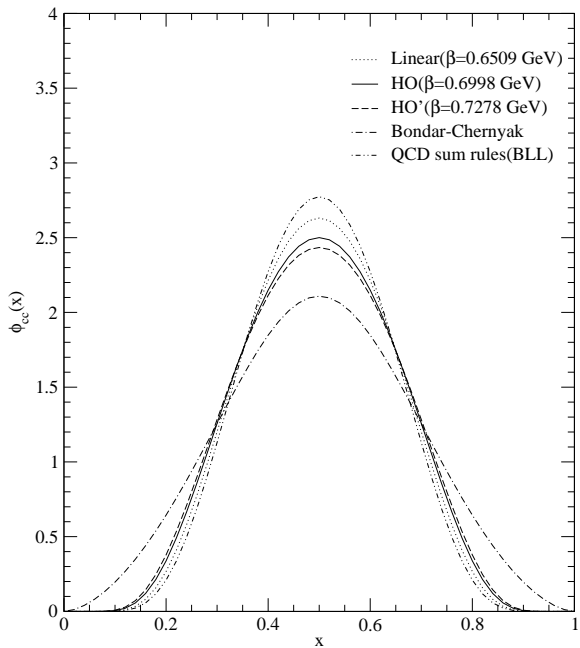}
\caption{The leading twist distribution amplitudes $\phi(x)$ for
$\eta_c$ and $J/\psi$[$\phi_{\eta_c}(x)\approx\phi_{J/\psi}(x)]$ obtained
from our LFQM compared with the ones from 
Bondar-Chernyak model~\cite{BC} and QCD sum rules~\cite{BLL}.}
\label{fig3}
\end{figure}
The shape of the quark DA which depends on ($m_c$, $\beta_{cc}$) values
is important to the calculation of the cross section for the heavy meson
pair production in $e^+e^-$ annihilations.
We thus show in Fig.~\ref{fig3} the normalized quark DA for $\eta_c$ and $J/\psi$,
$\phi_{c\bar{c}}(x)=\phi_{\eta_c}(x)\approx\phi_{J/\psi}(x)$ obtained from 
linear(dotted line), HO(solid line), and HO$^\prime$(dashed line) potentials
compared with the ones obtained from Bondar and Chernyak(BC)~\cite{BC}
(dot-dashed line) and from QCD sum rules~\cite{BLL}(doubledot-dashed line). 
As one can see from Fig.~\ref{fig3}, our quark DA $\phi_{c\bar{c}}(x)$ 
obtained at scale $\mu\simeq m_c$ practically vanishes in the
regions $x < 0.1$ and $x> 0.9$ where the motion of $c\bar{c}$ pair is expected
to be highly relativistic. 
However, our results
for quark DA are certaintly wider than the delta function-type(i.e. 
$\beta_{cc}\to 0$ limit) NRQCD results~\cite{BJ03,LHC03,HKQ03}, which do
not take into account the relative motion of valence quark-antiquark pair. 
Our results also show that the shape of quark DA becomes broader and more 
enhanced at the endpoint region($x\to0$ or 1) 
as the Gaussian parameter $\beta$(or equivalently transverse
${\bf k}_\perp$-size) increases. In comparison with other theoretical model 
calculations, we find that our result is quite 
consistent with the one obtained from QCD sum rules~\cite{BLL} at scale 
$\mu\simeq m_c$ but much narrower than the one obtained from BC~\cite{BC}. 
As will be discussed later, the cross section for double-charm production is
indeed very sensitive to the end point behavior of the quark DA.

\begin{table*}[t]
\caption{
The $\xi$ moments $\la\xi^n\ra_{\eta_c}$ and 
$\la\xi^n\ra_{J/\psi}=\la\xi^n\ra_L\approx\la\xi^n\ra_T$ for $\eta_c$ and  
$J/\psi$ distribution amplitudes obtained from our LFQM
at the scale $\mu\simeq m_c$ and compared with other 
model($\la\xi^n\ra_{\eta_c}\approx\la\xi^n\ra_{J/\psi}$)
estimates. Our central, upper, and
lower values are obtained from the HO, HO$^\prime$, and linear potential
parameters, respectively.}\label{t3}
\begin{tabular}{ccccccccc} \hline\hline
$\la\xi^n\ra$ & Ours& Ours & Buchmuller
& Cornell & BC~\cite{BC} & BKL~\cite{BKL} &NRQCD~\cite{BKL2} & QCD sum\\
& $\la\xi^n\ra_{\eta_c}$& $\la\xi^n\ra_{J/\psi}$ & Tye model~\cite{BT}
& model~\cite{Cor} & &  & & rules~\cite{BLL,Bra} \\
\hline
$n=2$ & 0.084$^{+0.004}_{-0.007}$ &0.082$^{+0.004}_{-0.006}$
&0.086 & 0.084 & 0.13 & 0.019 & $0.075\pm0.011$ & $0.070\pm0.007$ \\
\hline
$n=4$ & 0.017$^{+0.001}_{-0.003}$ &0.016$^{+0.002}_{-0.002}$
& 0.020 & 0.019 & 0.040 & 0.0083 &$0.010\pm0.003$ & $0.012\pm0.002$\\
\hline
$n=6$ &  0.0047$^{+0.0006}_{-0.0010}$ & 0.0046$^{+0.0005}_{-0.0010}$
&0.0066 &0.0066 &0.018 &0.0026 &$0.0017\pm0.0007$
& $0.0031\pm0.0008$\\
\hline\hline
\end{tabular}
\end{table*}
In Table~\ref{t3}, we list the calculated $\la\xi^n\ra$ moments up to $n=6$
for the $\eta_c$ and $J/\psi$ DAs at scale $\mu\simeq m_c$ and
compare with other model estimates. Our central, upper and lower values are
obtained from HO, HO$^\prime$ and linear parameters, respectively.
Since the $\xi$ moments for $J/\psi$ meson with the longitudinal 
polarization are almost the same as those with the transverse 
polarization, our results imply that 
$\la\xi^n\ra_{J/\psi}=\la\xi^n\ra_L=\la\xi^n\ra_T$, 
which is also confirmed by the recent QCD sum rule calculations~\cite{BLL,Bra}. 
Furthermore, the  $\xi$ moments between $\eta_c$ and $J/\psi$ mesons are not
much different from each other as one can see from Table~\ref{t3}.
Our results for the $\xi$ moments are in good agreement with those obtained
from other potential models~\cite{BT,Cor} as well as 
QCD sum-rules~\cite{BLL,Bra},
but disagree with the predictions obtained from BC~\cite{BC} and
BKL~\cite{BKL} models.
While NRQCD predictions~\cite{BKL2} for the second and fourth moments
are in agreement with our model but disagree for the higher moment 
$\la\xi^6\ra$. This disagreement for the moment
$\la\xi^6\ra$ may be ascribed to the end point behavior(i.e. relativistic
correction) of quark DA.

Within our model calculation, the relative quark velocity
can be obtained from the relation $m_rv^2/2\simeq 2\sqrt{m^2_c+{\vec k}^2}-2m_c$
in the center of mass frame, where $m_r$ is the reduced mass. From this
relation, we obtain $v^2\simeq 2{\bf k}_\perp^2/m^2_c=2\beta^2/m^2_c$,i.e.
\bea\label{v2LFQM}
\la v^2\ra_{c\bar{c}} = 0.30^{+0.02}_{-0.04},
\eea
where the central, upper, and lower values are from the HO,
HO$^\prime$, and linear potential parameters, respectively.
Using the dimensional regularization at leading order of $\al_s$, the 
authors in Refs.~\cite{BLL,Bra} derived the relation between 
the relative velocity of quark-antiquark pair inside the charmonium
and the $\xi$ moment as 
$\la\xi^n\ra_{\eta_c}\approx\la\xi^n\ra_{J/\psi}=\la v^n\ra/(n+1)(n=2,4,6)$. 
Applying this formula to our model calculations, we get the relative $v^2$ as
$\la v^2\ra_{c\bar{c}} = 0.25^{+0.01}_{-0.02}$,
where again the central, upper, and lower values are from the HO, 
HO$^\prime$, and linear potential parameters, respectively. 
The results obtained from our LFQM and QCD sum rule methods are 
not only in an agreement with each other but also quite consistent
with the value $\la v^2\ra_{c\bar{c}}\approx 0.3$ used in 
NRQCD~\cite{BJ03,BKL}. Note that one gets the
quark DA $\phi(x)\sim\delta(x-1/2)$ in the limit $v\to 0$ while 
$\phi(x)\sim\phi_{\rm as}(x)=6x(1-x)$ as $v\to 1$. As noted in QCD
sum rule calculations~\cite{BLL}, the moments $\la\xi^n\ra$ are proportional
to $v^n$ according to the NRQCD $v$-scaling rules~\cite{VS}:
$(m_cv^2)^2\ll(m_cv)^2\ll m^2_c$. 
It is not difficult to see that the $\xi$-moments
obtained from our LFQM and QCD sum rules~\cite{BLL} satisfy these
rules. 
However, as discussed in~\cite{BLL}, the BC moments~\cite{BC} break the
NRQCD $v$-scaling rules and 
the quark DA obtained from~\cite{BC} corresponds 
to the QCD sum rule result~\cite{BLL} defined at scale $\mu\simeq 10$ 
GeV rather than at $\mu\simeq m_c$. 
In our LFQM calculation, we would overestimate 
the experimental values of decay constants for $J/\psi$ and $\eta_c$ 
if we were to use the shape of BC distribution to get the cross section value
of double-charm production consistent with the experimental data.
Thus, it seems misleading to claim that 
the cross section of $e^+e^-\to J/\psi+\eta_c$ 
in~\cite{BC} led to a good agreement with the experiment.

The corresponding Gegenbauer moments obtained from Eq.~(\ref{xia}) 
are given by
\bea\label{Geta}
a_2(\mu\simeq m_c)&=&-0.339^{+0.011}_{-0.021},
\nonumber\\
a_4(\mu\simeq m_c)&=&0.082^{-0.010}_{+0.022},
\nonumber\\
a_6(\mu\simeq m_c)&=&+0.0027^{+0.0041}_{-0.0112},
\eea
for $\eta_c$ meson and
\bea\label{GJpsi}
a_2(\mu\simeq m_c)&=&-0.343^{+0.011}_{-0.020},
\nonumber\\
a_4(\mu\simeq m_c)&=&0.087^{-0.010}_{+0.020},
\nonumber\\
a_6(\mu\simeq m_c)&=&-0.0015^{+0.0035}_{-0.0067},
\eea
for $J/\psi$ meson, respectively. Since $a_n$ for $J/\psi$ with
longitudinal polarization and $a^T_n$ with transverse polarization are
not much different from each other, we do not distinguish them in our model calculation.

\begin{figure*}[t]
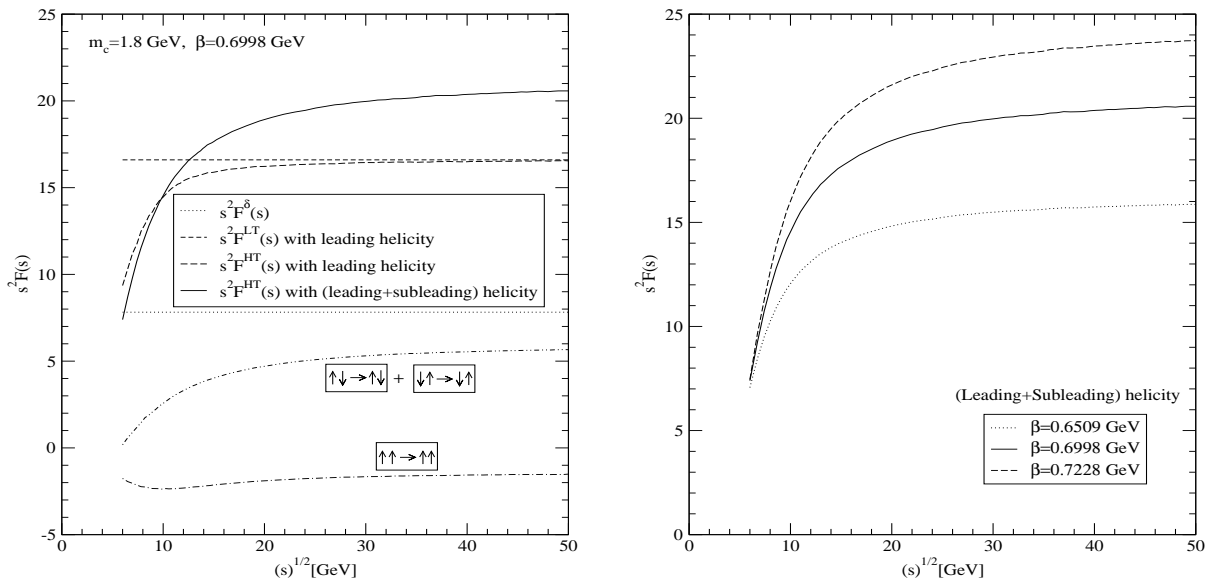

\vspace{1.1cm}
\includegraphics[width=3.0in,height=3.0in]{Fig4a}
\hspace{0.5cm}
\includegraphics[width=3.0in,height=3.0in]{Fig4b}
\caption{The form factor $s^2{\cal F}(s)$ for $e^+e^-\to J/\psi+\eta_c$.
The dotted, short-dashed, long-dashed and solid line in left panel represent 
the peaking, leading twist(LT), higher twist(HT) results with the leading helicty
contributions and HT one with all helicity contributions, respectively.
The dot-dashed and double-dot-dashed line represent the dominant subleading
contributions. The right panel represents the HT results including all helicity
contributions obtained from linear(dotted line), HO(solid line), and HO'(dashed line)
model parameters, respectively.}
\label{fig4}
\end{figure*}
In Fig.~\ref{fig4}, we show $s^2{\cal F}(s)$ for $e^+e^-\to J/\psi+\eta_c$ 
process.  The left panel of Fig.~\ref{fig4} shows the results 
obtained from the central value $\beta=0.6998$ GeV of our model parameters
displaying different(leading and subleading) helicity contributions. The 
right panel of Fig.~\ref{fig4} shows the sensitivity of our model 
predictions with all helicity contributions when
the gaussian model parameter $\beta$ changes as shown in Fig.~\ref{fig3}.
In the left panel, the dotted and short-dashed lines represent
the results obtained from the non-relativistic peaking approximation
${\cal F}^\delta(s)$[ Eq.~(\ref{T_peak})]
and the leading twist(LT) factorized form factor ${\cal F}^{\rm LT}(s)$[
Eq.~(\ref{F_fac})] taking into account the relative motion of valence quarks, 
respectively. The long-dashed line represents the higher twist(HT) 
nonfactorized form factor ${\cal F}^{\rm HT}(s)$[Eq.~(\ref{FNR})] obtained 
by including the transverse momenta$({\bf k}_\perp,{\bf l}_\perp)$ both 
in the wave function and the hard scattering part. 
Note that ${\cal F}^\delta(s)$(dotted line), 
${\cal F}^{\rm LT}(s)$(short-dashed line), and 
${\cal F}^{\rm HT}(s)$(long-dashed line) are obtained from the leading
helicity contributions. The solid line represents our full solution 
${\cal F}^{\rm HT}_{\rm (\Delta H=0 + \Delta H=1)}(s)$
including all(leading plus subleading) helicity contributions summarized in
the Appendix A(Tables~\ref{t4} and ~\ref{t5}).
Among the subleading helicity contributions,  we find that only 
$\uparrow\uparrow\to\uparrow\uparrow$(dot-dashed line) and 
$(\uparrow\downarrow\to\uparrow\downarrow)+
(\downarrow\uparrow\to\downarrow\uparrow)$(double-dot-dashed line) 
helicity contributions give a sizeable effects  and 
other subleading helicity contributions are negligible. 
As shown in Fig.~\ref{fig4}, we find that while
$s^2{\cal F}^{\rm HT}_{\rm (\Delta H=0 + \Delta H=1)}(s)$
is about 2 times larger than $s^2{\cal F}^\delta(s)$  
but 10$\%$ smaller than $s^2{\cal F}^{\rm LT}(s)$ at $\sqrt{s}=10.6$ GeV.
It is also interesting
to note that our $s^2{\cal F}^{\rm HT}_{\rm (\Delta H=0 + \Delta H=1)}(s)$
takes over $s^2{\cal F}^{\rm LT}(s)$ 
for $\sqrt{s}\gtrsim 13$ GeV region, although 
$s^2{\cal F}^{\rm HT}(s)$ with leading helicity components 
approaches to $s^2{\cal F}^{\rm LT}(s)$ as $s\to\infty$. 
As one can see from Fig.~\ref{fig4}, the form factor obtained from
our calculation shows ${\cal F}(s)\sim s^{-2}$ as $s\to\infty$
which is the expected QCD scaling behavior\cite{Ma,JMY,BF,MMT,CJ,BJ}
for the transition form factor between pseudoscalar ($0^{-+}$)
and vector ($1^{--}$) mesons. 

\begin{figure*}[t]
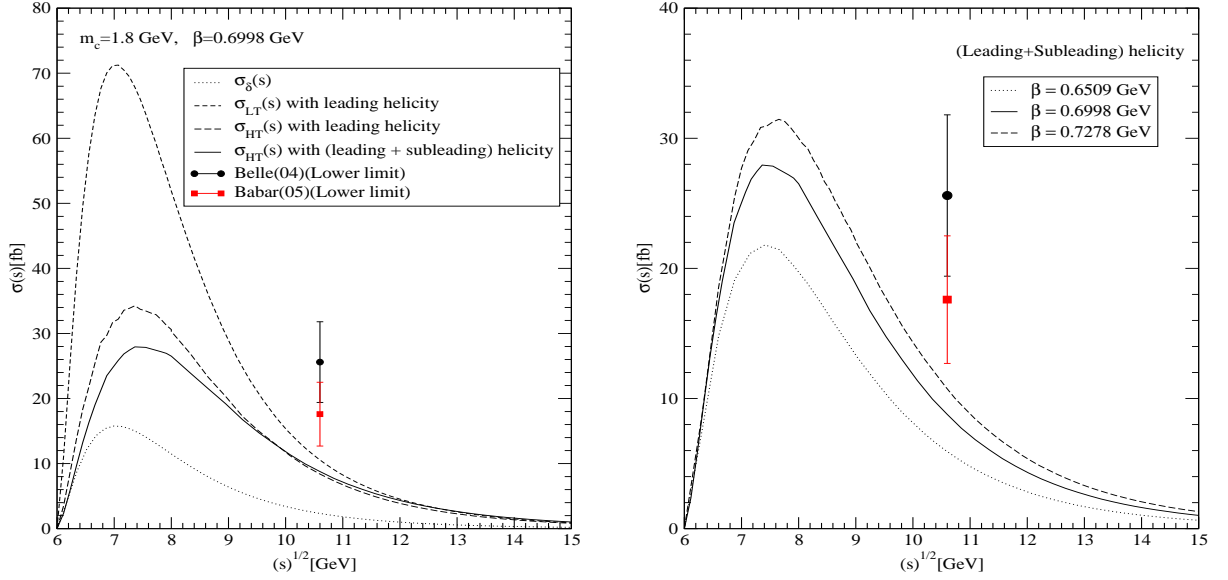

\vspace{1.1cm}
\includegraphics[width=3.0in,height=3.0in]{Fig5a}
\hspace{0.5cm}
\includegraphics[width=3.0in,height=3.0in]{Fig5b}
\caption{The cross section for $e^+e^-\to J/\psi\eta_c$ with leading and subleading
helicity contributions using the HO model parameters(left panel) and with all
helicity contributions using the linear, HO, and HO' model parameters(right panel).}
\label{fig5}
\end{figure*}
In Fig.5, we show leading order in $\al_s$ contribution to
the cross section for $e^+e^-\to J/\psi+\eta_c$.
The left panel of Fig.~\ref{fig5} shows the results  with leading and subleading
helicity contributions using the HO model parameters.
The right panel of  Fig.~\ref{fig5} shows the sensitivity of our model predictions 
with all helicity contributions when
the gaussian model parameter $\beta$ changes as shown in Figs.~\ref{fig3} and ~\ref{fig4}.
The line codes are the same as in Fig.~\ref{fig4}.
As one can see from  the left panel of Fig.5, 
our peaking approximation 
result(dotted line) is consistent with the previous NRQCD estimates in 
Refs.~\cite{BJ03,LHC03,HKQ03}, which is an order of magnitude smaller than 
the experimental data~\cite{Belle,Babar}. 
We should note from the left panel of of Fig.5 that
our higher twist result(solid line) including all helicity contributions
enhances the peaking approximation result by a factor of 
$3\sim 4$ at $\sqrt{s}=10.6$ GeV while it reduces that of the leading twist  
result by $20\%$.
As discussed in Section III, our higher twist results ($\sigma_{\rm HT}$)
include all orders of 
${\bf k}^2_\perp/{\bf q}^2_\perp$, ${\bf l}^2_\perp/{\bf q}^2_\perp$ 
and ${\bf k}_\perp\cdot{\bf l}_\perp/{\bf q}^2_\perp$
to keep effectively all higher orders of the relative quark velocity
beyond $\la v^2\ra$. If we keep only the leading order of these terms
(${\bf k}^2_\perp/{\bf q}^2_\perp$, ${\bf l}^2_\perp/{\bf q}^2_\perp$ 
and ${\bf k}_\perp\cdot{\bf l}_\perp/{\bf q}^2_\perp$),
our results would correspond to include the relativistic effects up
to the order of $\la v^2\ra$\cite{EM}.
Our predictions for the cross section at $\sqrt{s}=10.6$ GeV obtained from
peaking approximation($\sigma_{\delta}$), 
leading twist($\sigma_{\rm LT}$) and
higher twist($\sigma_{\rm HT}$) are given by
\bea\label{CSNR}
\sigma_{\delta}(J/\psi+\eta_c)
&=&2.34^{+0.50}_{-0.69}[\rm fb],
\nonumber\\
\sigma_{\rm LT}(J/\psi+\eta_c)
&=&10.57^{+3.15}_{-4.02}[\rm fb],
\nonumber\\
\sigma^{(\Delta H=0+\Delta H=1)}_{\rm HT}(J/\psi+\eta_c)
&=&8.76^{+1.61}_{-2.84}[\rm fb],
\eea
where the central, upper and lower values are obtained from HO, HO$^\prime$ and
linear potential parameters, respectively. 
Our prediction of $\sigma^{(\Delta H=0+\Delta H=1)}_{\rm HT}(J/\psi+\eta_c)$
reduces by about $10\%$ from the value in Eq.(\ref{CSNR}) to
$7.68^{+1.94}_{-2.66}$[fb] when we keep only the leading order of 
${\bf k}^2_\perp/{\bf q}^2_\perp$, ${\bf l}^2_\perp/{\bf q}^2_\perp$ 
and ${\bf k}_\perp\cdot{\bf l}_\perp/{\bf q}^2_\perp$.
It is interesting to note that our reduced value 
$7.68^{+1.94}_{-2.66}$[fb] is indeed
very close to the result $7.8 [\rm fb]$ obtained in the recent 
investigation including the relativistic effects up to $\la v^2\ra$\cite{EM}.

\begin{figure}[t]
\vspace{1.1cm}
\includegraphics[width=3.0in,height=3.0in]{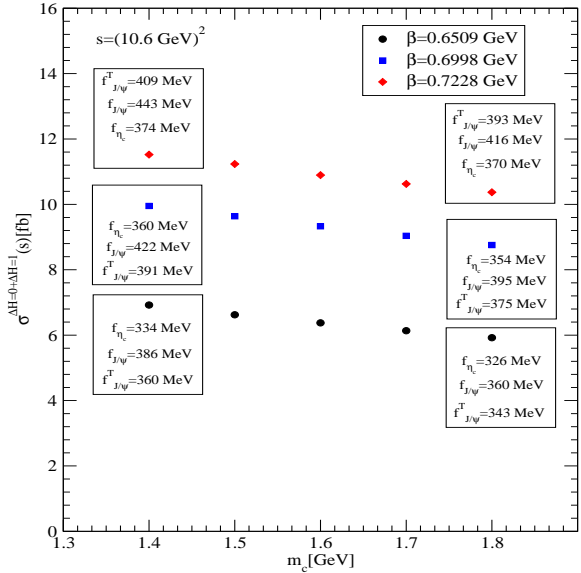}
\caption{The parameter ($m_c,\beta$) dependence of the cross section
for $e^+e^-\to J/\psi+\eta_c$.}
\label{fig6}
\end{figure}

As a sensitivity check, we show in Fig.~\ref{fig6} the parameter ($m_c,\beta$) 
dependence of the cross section for $e^+e^-\to J/\psi+\eta_c$
using the nonfactorized higher twist form factor with all helicty 
contributions. We also show in  Fig.~\ref{fig6}
the decay constants corresponding to the end point mass values, 
$m_c=1.4$ GeV and 1.8 GeV.
The cross section increases as $\beta(m_c)$ increases(decreases).  
As one can also see from Fig.~\ref{fig6}, the cross 
section is more sensitive to the variation of the gaussian parameter than 
to the variation of the charm quark mass.

The experimental results are 
\be\label{CS_Belle}
\sigma(J/\psi+\eta_c)\times B^{\eta_c}[\geq 2]
=(25.6\pm2.8\pm3.4)[\rm fb],
\ee
by Belle~\cite{Belle}(filled circle in Fig~\ref{fig5}) and
\be\label{CS_Babar}
\sigma(J/\psi+\eta_c)\times B^{\eta_c}[\geq 2]
=(17.6\pm2.8^{+1.5}_{-2.1})[\rm fb],
\ee
by Babar~\cite{Babar}(filled square in Fig~\ref{fig5}), 
where $B^{\eta_c}[\geq 2]$ is the branching 
fraction for $\eta_c$ decay into at least two charged particles.
Considering an enhancement by the factor of $1.8$ from the corrections of
next-to-leading order (NLO) of $\al_s$\cite{ZGC}, it might be conceivable
to raise our leading $\al_s$ order result 
$\sigma^{(\Delta H=0+\Delta H=1)}_{\rm HT}(J/\psi+\eta_c)$
in Eq.(\ref{CSNR}) by this factor and get 
a value close to the above Babar data.
However, it would be necessary to make 
detailed NLO investigation within the LF PQCD framework 
before we can make any firm conclusion.


\section{Summary and Conclusion}
We investigated the transverse momentum effect on the exclusive charmonium
$J/\psi+\eta_c$ pair production in $e^+e^-$ annihilation using the 
nonfactorized PQCD and LFQM that goes beyond the peaking approximation. 

Our LFQM calculation based on the variational principle for the 
QCD-motivated Hamiltonian~\cite{CJ1,CJ2} shows that 
the quark DAs for $J/\psi$ and $\eta_c$ take substantially broad shape 
which is quite different from the $\delta$-type DA. 
If the quark DA is not an exact
$\delta$ function, i.e. the relative motion of valence quarks can play a 
significant role, the factorization theorem is no longer applicable.  
In going beyond the peaking approximation, we stressed a consistency
by keeping the transverse momentum ${\bf k}_\perp$ both in the wave function 
part and the hard scattering part simultaneously before doing any integration
in the amplitude. Such non-factorized analysis should be distinguished from 
the factorized analysis where the transverse momenta are seperately integrated 
out in the wave function part and in the hard scattering part.
Even if the used LF wave functions lead to the similar shapes of DAs,
predictions for the cross sections of double-charm productions are 
apparently different between the factorized and non-factorized analyses.
We found that the higher twist contributions including all 
helicity contributions enhanced NRQCD result by a factor of
$3\sim 4$ at $\sqrt{s}=10.6$ GeV while it reduced that of the leading twist
result by $20\%$. We also found that the cross
section for $e^+e^-\to J/\psi\eta_c$ process at $\sqrt{s}=10.6$ GeV
is more sensitive to the variation of the gaussian parameter than
to that of the charm quark mass. Our results showed that
the cross section increases as $\beta(m_c)$ increases(decreases). 

In conclusion, LFQM/PQCD analysis showed that the relativistic 
correction(i.e. non-delta function) of the light-front wave function 
is very important to understand the large discrepancy between 
the NRQCD result and the 
experimental data given by Eqs.(\ref{CS_Belle}) and (\ref{CS_Babar}). 
While there have been 
considerations of broadening the quark DA to reduce the discrepancy 
between the theory at the leading order of $\al_s$ and the experimental 
results\cite{BC,Ma,BLL05}, a recent calculation of corrections of 
next-to-leading order(NLO) of $\al_s$ leads to an enhancement of
the theoretical prediction by the factor about 1.8~\cite{ZGC}. 
This factor may enhance our result in the leading order of $\al_s$ to
fit the current experimental results. However, more detailed investigation
is necessary prior to any firm conclusion on this issue. 

\acknowledgments
This work was supported by a grant from the U.S. Department of
Energy(No. DE-FG02-03ER41260).  H.-M. Choi was supported in part by Korea
Research Foundation under the contract KRF-2005-070-C00039. The National
Energy Research Scientific Center is also acknowledged for the grant of
supercomputing time.

\appendix
\section{Helicity contributions to the hard scattering amplitude}
In this appenix A, we summarize the helicity contributions
$(\lam_1,\lam_2)\to(\lam'_1,\lam'_2)$ to the hard scattering amplitude
$T^{(\lam_1,\lam_2)\to(\lam'_1,\lam'_2)}_H$ for the 
$\eta_c(P)\to\gamma^*(q) + J/\psi(P')$ process.

\begin{table*}[t]
\caption{Helicity contributions to the hard scattering amplitude $T_A$
in Fig.~\ref{fig2}.}\label{t4}
\begin{tabular}{cccccc} \hline\hline
$|\Delta H|$ & Helicities & $N_{A_1}$ & $N_{A_2}$ & $N_{A_3}$ 
&$\sum_iT_{A_i}(\Delta_x=\Delta_y=0)$\\
\hline
 & $\up\down\to\up\down$ & $P_{A1}+iQ_A$ & $P_{A2}+iQ_A$ & $P_{A3}+iQ_A$ 
& $-\frac{8\pi\al_s C_F}{(y_2-x_2)}[\frac{P_A+iQ}{{\cal D}_1{\cal D}_2}
+ \frac{2}{(y_2-x_2){\cal D}_2}]$\\
 & $\down\up\to\down\up$ & $P_{A1}-iQ_A$ & $P_{A2}-iQ_A$ & $P_{A3}-iQ_A$ 
& $-\frac{8\pi\al_s C_F}{(y_2-x_2)}[\frac{P_A-iQ}{{\cal D}_1{\cal D}_2}
+ \frac{2}{(y_2-x_2){\cal D}_2}]$\\
0 & $\up\down\to\down\up$ & $-\frac{2m^2(y_1-x_1)(y_2-x_2)}{x_1x_2y_1y_2}$ 
& $N_{A_1}^{(\up\down\to\down\up)}$ & $N_{A_1}^{(\up\down\to\down\up)}$ 
& $-\frac{8\pi\al_s C_F}{(y_2-x_2)
{\cal D}_1{\cal D}_2}N_{A_1}^{(\up\down\to\down\up)}$\\
 & $\down\up\to\up\down$ & $-\frac{2m^2(y_1-x_1)(y_2-x_2)}{x_1x_2y_1y_2}$ 
& $N_{A_1}^{(\down\up\to\up\down)}$ & $N_{A_1}^{(\down\up\to\up\down)}$ 
& $-\frac{8\pi\al_s C_F}{(y_2-x_2)
{\cal D}_1{\cal D}_2}N_{A_1}^{(\down\up\to\up\down)}$\\
 & $\up\up\to\up\up$ & $F_{A1}+iG_A$ &  $F_{A2}+iG_A$ & $F_{A3}+iG_A$ 
& $-\frac{8\pi\al_s C_F}{(y_2-x_2)}[\frac{F_A+iG}{{\cal D}_1{\cal D}_2}
+ \frac{2}{(y_2-x_2){\cal D}_2}]$\\
 & $\down\down\to\down\down$ & $F_{A1}-iG_A$ &  $F_{A2}-iG_A$ &  $F_{A3}-iG_A$ 
& $-\frac{8\pi\al_s C_F}{(y_2-x_2)}[\frac{F_A-iG}{{\cal D}_1{\cal D}_2}
+ \frac{2}{(y_2-x_2){\cal D}_2}]$\\
\hline
 &$\up\down\to\up\up$ 
&  $\frac{2m(x_1 l - y_1 k - x_2y_1 q)^L}{x_2y_1y_2}$
& $N_{A_1}^{(\up\down\to\up\up)}$ & $N_{A_1}^{(\up\down\to\up\up)}$ 
& $-\frac{8\pi\al_s C_F}{(y_2-x_2)
{\cal D}_1{\cal D}_2}N_{A_1}^{(\up\down\to\up\up)}$\\
&$\up\down\to\down\down$ 
& $\frac{2m(x_2 l - y_2 k - x_2y_2 q)^R}{x_1y_1y_2}$ 
& $N_{A_1}^{(\up\down\to\down\down)}$ & $N_{A_1}^{(\up\down\to\down\down)}$ 
& $-\frac{8\pi\al_s C_F}{(y_2-x_2)
{\cal D}_1{\cal D}_2}N_{A_1}^{(\up\down\to\down\down)}$\\
 &$\down\up\to\up\up$ 
& $-\frac{2m(x_2 l - y_2 k - x_2y_2 q)^L}{x_1y_1y_2}$ 
& $N_{A_1}^{(\down\up\to\up\up)}$ & $N_{A_1}^{(\down\up\to\up\up)}$  
& $-\frac{8\pi\al_s C_F}{(y_2-x_2)
{\cal D}_1{\cal D}_2}N_{A_1}^{(\down\up\to\up\up)}$\\
1 &$\down\up\to\down\down$ 
& $-\frac{2m(x_1 l - y_1 k - x_2y_1 q)^R}{x_2y_1y_2}$ 
& $N_{A_1}^{(\down\up\to\down\down)}$ & $N_{A_1}^{(\down\up\to\down\down)}$ 
& $-\frac{8\pi\al_s C_F}{(y_2-x_2)
{\cal D}_1{\cal D}_2}N_{A_1}^{(\down\up\to\down\down)}$\\
 &$\up\up\to\up\down$ 
& $-\frac{2m(x_1 l - y_1 k - x_2y_1 q)^R}{x_1x_2y_2}$ 
& $N_{A_1}^{(\up\up\to\up\down)}$ & $N_{A_1}^{(\up\up\to\up\down)}$ 
& $-\frac{8\pi\al_s C_F}{(y_2-x_2)
{\cal D}_1{\cal D}_2}N_{A_1}^{(\up\up\to\up\down)}$\\
 &$\down\down\to\up\down$ 
& $-\frac{2m(x_2 l - y_2 k - x_2y_2 q)^L}{x_1x_2y_1}$ 
& $N_{A_1}^{(\down\down\to\up\down)}$ & $N_{A_1}^{(\down\down\to\up\down)}$ 
& $-\frac{8\pi\al_s C_F}{(y_2-x_2)
{\cal D}_1{\cal D}_2}N_{A_1}^{(\down\down\to\up\down)}$\\
 &$\up\up\to\down\up$ 
& $\frac{2m(x_2 l - y_2 k - x_2y_2 q)^R}{x_1x_2y_1}$ 
& $N_{A_1}^{(\up\up\to\down\up)}$ & $N_{A_1}^{(\up\up\to\down\up)}$ 
& $-\frac{8\pi\al_s C_F}{(y_2-x_2)
{\cal D}_1{\cal D}_2}N_{A_1}^{(\up\up\to\down\up)}$\\
 &$\down\down\to\down\up$ 
& $\frac{2m(x_1 l - y_1 k - x_2y_1 q)^L}{x_1x_2y_2}$ 
& $N_{A_1}^{(\down\down\to\down\up)}$ & $N_{A_1}^{(\down\down\to\down\up)}$ 
& $-\frac{8\pi\al_s C_F}{(y_2-x_2)
{\cal D}_1{\cal D}_2}N_{A_1}^{(\down\down\to\down\up)}$\\
\hline
2 & $\up\up\to\down\down$ & 0 & 0 & 0 & 0\\
 & $\down\down\to\up\up$ & 0 & 0 & 0  & 0\\
\hline\hline
\end{tabular}
\end{table*}
\begin{table*}[t]
\caption{Helicity contributions to the hard scattering amplitude $T_B$
in Fig.~\ref{fig2}.}\label{t5}
\begin{tabular}{cccccc} \hline\hline
$|\Delta H|$ & Helicities & $N_{B_1}$ & $N_{B_2}$ & $N_{B_3}$ 
& $\sum_iT_{B_i}(\Delta_x=\Delta_y=0)$\\
\hline
 & $\up\down\to\up\down$ & $P_{B1}+iQ_B$ & $P_{B2}+iQ_B$ & $P_{B3}+iQ_B$ 
& $-\frac{8\pi\al_s C_F}{(y_2-x_2)}[\frac{P_B+iQ_B}{{\cal D}_2{\cal D}_8}
+ \frac{2}{(y_2-x_2){\cal D}_5}]$\\
 & $\down\up\to\down\up$ & $P_{B1}-iQ_B$ & $P_{B2}-iQ_B$ & $P_{B3}-iQ_B$ 
& $-\frac{8\pi\al_s C_F}{(y_2-x_2)}[\frac{P_B-iQ_B}{{\cal D}_2{\cal D}_8}
+ \frac{2}{(y_2-x_2){\cal D}_5}]$ \\
0 & $\up\down\to\down\up$ & $-\frac{2m^2(y_1-x_1)(y_2-x_2)}{x_1x_2y_1y_2}$
& $N_{B_1}^{(\up\down\to\down\up)}$ & $N_{B_1}^{(\up\down\to\down\up)}$ 
& $-\frac{8\pi\al_s C_F}{(y_2-x_2)
{\cal D}_2{\cal D}_8}N_{B_1}^{(\up\down\to\down\up)}$ \\
 & $\down\up\to\up\down$ & $-\frac{2m^2(y_1-x_1)(y_2-x_2)}{x_1x_2y_1y_2}$
& $N_{B_1}^{(\down\up\to\up\down)}$ & $N_{B_1}^{(\down\up\to\up\down)}$ 
& $-\frac{8\pi\al_s C_F}{(y_2-x_2)
{\cal D}_2{\cal D}_8}N_{B_1}^{(\down\up\to\up\down)}$\\
& $\up\up\to\up\up$ & $F_{B1}+iG_B$ &  $F_{B2}+iG_B$ & $F_{B3}+iG_B$ 
& $-\frac{8\pi\al_s C_F}{(y_2-x_2)}[\frac{F_B+iG_B}{{\cal D}_2{\cal D}_8}
+ \frac{2}{(y_2-x_2){\cal D}_5}]$ \\
& $\down\down\to\down\down$ & $F_{B1}-iG_B$ & $F_{B2}-iG_B$ & $F_{B3}-iG_B$ 
& $-\frac{8\pi\al_s C_F}{(y_2-x_2)}[\frac{F_B-iG_B}{{\cal D}_2{\cal D}_8}
+ \frac{2}{(y_2-x_2){\cal D}_5}]$ \\
\hline
&$\up\down\to\up\up$
&  $\frac{2m(x_1 l - y_1 k - x_1y_2 q)^L}{x_2y_1y_2}$
& $N_{B_1}^{(\up\down\to\up\up)}$ & $N_{B_1}^{(\up\down\to\up\up)}$ 
& $-\frac{8\pi\al_s C_F}{(y_2-x_2)
{\cal D}_2{\cal D}_8}N_{B_1}^{(\up\down\to\up\up)}$\\
&$\up\down\to\down\down$
& $\frac{2m(x_2 l - y_2 k - x_2y_2 q)^R}{x_1y_1y_2}$
& $N_{B_1}^{(\up\down\to\down\down)}$ & $N_{B_1}^{(\up\down\to\down\down)}$ 
& $-\frac{8\pi\al_s C_F}{(y_2-x_2)
{\cal D}_2{\cal D}_8}N_{B_1}^{(\up\down\to\down\down)}$\\
 &$\down\up\to\up\up$
& $-\frac{2m(x_2 l - y_2 k - x_2y_2 q)^L}{x_1y_1y_2}$
& $N_{B_1}^{(\down\up\to\up\up)}$ & $N_{B_1}^{(\down\up\to\up\up)}$  
& $-\frac{8\pi\al_s C_F}{(y_2-x_2)
{\cal D}_2{\cal D}_8}N_{B_1}^{(\down\up\to\up\up)}$\\
1 &$\down\up\to\down\down$
& $-\frac{2m(x_1 l - y_1 k - x_1y_2 q)^R}{x_2y_1y_2}$
& $N_{B_1}^{(\down\up\to\down\down)}$ & $N_{B_1}^{(\down\up\to\down\down)}$ 
& $-\frac{8\pi\al_s C_F}{(y_2-x_2)
{\cal D}_2{\cal D}_8}N_{B_1}^{(\down\up\to\down\down)}$\\
 &$\up\up\to\up\down$
& $-\frac{2m(x_1 l - y_1 k - x_1y_2 q)^R}{x_1x_2y_2}$
& $N_{B_1}^{(\up\up\to\up\down)}$ & $N_{B_1}^{(\up\up\to\up\down)}$ 
& $-\frac{8\pi\al_s C_F}{(y_2-x_2)
{\cal D}_2{\cal D}_8}N_{B_1}^{(\up\up\to\up\down)}$\\
 &$\down\down\to\up\down$
& $-\frac{2m(x_2 l - y_2 k - x_2y_2 q)^L}{x_1x_2y_1}$
& $N_{B_1}^{(\down\down\to\up\down)}$ & $N_{B_1}^{(\down\down\to\up\down)}$ 
& $-\frac{8\pi\al_s C_F}{(y_2-x_2)
{\cal D}_2{\cal D}_8}N_{B_1}^{(\down\down\to\up\down)}$\\
 &$\up\up\to\down\up$
& $\frac{2m(x_2 l - y_2 k - x_2y_2 q)^R}{x_1x_2y_1}$
& $N_{B_1}^{(\up\up\to\down\up)}$ & $N_{B_1}^{(\up\up\to\down\up)}$ 
& $-\frac{8\pi\al_s C_F}{(y_2-x_2)
{\cal D}_2{\cal D}_8}N_{B_1}^{(\up\up\to\down\up)}$\\
 &$\down\down\to\down\up$
& $\frac{2m(x_1 l - y_1 k - x_1y_2 q)^L}{x_1x_2y_2}$
& $N_{B_1}^{(\down\down\to\down\up)}$ & $N_{B_1}^{(\down\down\to\down\up)}$ 
& $-\frac{8\pi\al_s C_F}{(y_2-x_2)
{\cal D}_2{\cal D}_8}N_{B_1}^{(\down\down\to\down\up)}$\\
\hline
2 & $\up\up\to\down\down$ & 0 & 0 & 0 & 0\\
 & $\down\down\to\up\up$ & 0 & 0 & 0 & 0\\
\hline\hline
\end{tabular}
\end{table*}
In Tables~\ref{t4} and~\ref{t5}, we summarize our results for the helicity
contributions to the hard scattering amplitudes
$T_{A}$ and $T_{B}$ for the diagrams in Fig.~\ref{fig2}, where
\begin{widetext}
\bea\label{A1}
P_{A1}&=&
\frac{-2}{x_1x_2y_1y_2}\biggl[
x^2_2y_1y_2 {\bf q}^2_\perp + x_1x_2 {\bf l}^2_\perp
+ y_1y_2 {\bf k}^2_\perp
+ x_2(x_1y_1 + x_2y_2)({\bf l}_\perp\cdot {\bf q}_\perp)
\nonumber\\
&&+ 2x_2y_1y_2({\bf k}_\perp\cdot {\bf q}_\perp)
+ (x_1y_1 + x_2y_2)(m^2 + {\bf k}_\perp\cdot {\bf l}_\perp)
\biggr]
+
\frac{2}{(x_1-y_1)}[D_2 + D_4]
= P_A + \frac{2}{(x_1-y_1)}[D_2 + D_4],
\nonumber\\
P_{A2}&=& P_A + \frac{2}{(x_1-y_1)}[D_2 + D_4],\;\;
P_{A3}= P_A + \frac{2}{(x_1-y_1)}[D_2 - D_4],
\nonumber\\
Q_A&=&\frac{2}{x_1x_2y_1y_2}\biggl[
x_2(x_1 - y_2)|{\bf l}_\perp\times {\bf q}_\perp|
- 2x_1y_1y_2|{\bf k}_\perp\times {\bf q}_\perp|
+ (x_2y_2 + x_1y_1)|{\bf k}_\perp\times {\bf l}_\perp| \biggl],
\nonumber\\
F_{A1}&=& \frac{-2}{x_1x_2y_1y_2}\biggl[
{\bf k}_\perp\cdot {\bf l}_\perp + x_2({\bf l}_\perp\cdot {\bf q}_\perp)
+ (x_1y_1 + x_2y_2)m^2
\biggr] + \frac{2}{(x_1-y_1)}[D_2 + D_4]
= F_A + \frac{2}{(x_1-y_1)}[D_2 + D_4],
\nonumber\\
F_{A2}&=& F_A + \frac{2}{(x_1-y_1)}[D_2 + D_4],\;\;
F_{A3}= F_A + \frac{2}{(x_1-y_1)}[D_2 - D_4],
\nonumber\\
G_A&=&\frac{2}{x_1x_2y_1y_2}\biggl[
2x_1 y_1 y_2|{\bf k}_\perp\times {\bf q}_\perp|
- x_2|{\bf l}_\perp\times {\bf q}_\perp|
+ (1-2x_1y_1) |{\bf k}_\perp\times {\bf l}_\perp|
\biggr],
\eea
for the diagrams $A_i$ and 
\bea\label{A3}
P_{B1}&=&
\frac{-2}{x_1x_2y_1y_2}\biggl[
x_1x_2y^2_2 {\bf q}^2_\perp + x_1x_2 {\bf l}^2_\perp 
+ y_1y_2 {\bf k}^2_\perp
- y_2(x_1y_1 + x_2y_2)({\bf k}_\perp\cdot {\bf q}_\perp)
\nonumber\\
&&- 2x_1x_2y_2({\bf l}_\perp\cdot {\bf q}_\perp) 
+ (x_1y_1 + x_2y_2)(m^2 + {\bf k}_\perp\cdot {\bf l}_\perp)
\biggr]
+
\frac{2}{(y_2-x_2)}[D_9 + D_7]
= P_B + \frac{2}{(y_2-x_2)}[D_9 + D_7],
\nonumber\\
P_{B2}&=& P_B + \frac{2}{(x_2-y_2)}[D_9 + D_7],\;\;
P_{B3}= P_B+ \frac{2}{(y_2-x_2)}[D_9 - D_7],
\nonumber\\
Q_B&=&\frac{2}{x_1x_2y_1y_2}(x_1y_1-x_2y_2)
(|{\bf l}_\perp\times {\bf k}_\perp| 
- y_2 |{\bf q}_\perp\times {\bf k}_\perp|),
\nonumber\\
F_{B1}&=& \frac{-2}{x_1x_2y_1y_2}\biggl[
{\bf k}_\perp\cdot {\bf l}_\perp - y_2({\bf k}_\perp\cdot {\bf q}_\perp)
+ (x_1y_1 + x_2y_2)m^2
\biggr] + \frac{2}{(y_2-x_2)}[D_9 + D_7]
=F_B + \frac{2}{(y_2-x_2)}[D_9 + D_7],
\nonumber\\
F_{B2}&=& F_B + \frac{2}{(x_2-y_2)}[D_9 + D_7],\;\;
F_{B3}= F_B + \frac{2}{(y_2-x_2)}[D_9 - D_7],
\nonumber\\
G_B&=&\frac{2}{x_1x_2y_1y_2}
(y_2|{\bf q}_\perp\times {\bf k}_\perp| 
- |{\bf l}_\perp\times {\bf k}_\perp|),
\eea
for the diagrams $B_i$, respectively.
\end{widetext}

As an illustration, we show how to obtain the hard scattering amplitudes
$T_{A}=\sum^3_{i=1}T_{A_i}$ (sixth column in Table~\ref{t4}) and
$T_{B}=\sum^3_{i=1}T_{B_i}$(sixth column in Table~\ref{t5}) as well as
the total amplitude $T_H=T_A + T_B$
for the $(\up\down\to\up\down)$ contribution. Using the identities 
Eqs.~(\ref{DsumA}) and~(\ref{DsumB}) in Sec.III, we obtain
\begin{widetext}
\bea\label{A5}
\biggl[\sum_iT_{A_i}^{(\up\down\to\up\down)}\biggr]_{\Delta=0}
&=&-8\pi\al_s C_F\biggl[
\frac{\theta(y_2-x_2)(P_{A1} + iQ_A)}{(y_2-x_2){\cal D}_1{\cal D}_2}
+\frac{\theta(x_2-y_2)(P_{A2}+iQ_A)}{(x_2-y_2){\cal D}_3{\cal D}_4}
+\frac{\theta(x_2-y_2)(P_{A3} + iQ_A)}{(x_2-y_2){\cal D}_5{\cal D}_6}
\biggr]
\nonumber\\
&=&-8\pi\al_s C_F(P_A+iQ_A)\biggl[
\frac{\theta(y_2-x_2)}{(y_2-x_2){\cal D}_1{\cal D}_2}
+\frac{\theta(x_2-y_2)}{(x_2-y_2)}
\biggl(\frac{1}{{\cal D}_3{\cal D}_4}+\frac{1}{{\cal D}_5{\cal D}_6}\biggr)
\biggr]
\nonumber\\
&&-16\pi\al_s C_F\biggl[
\frac{\theta(y_2-x_2)}{(y_2-x_2)^2}\frac{{\cal D}_2+{\cal D}_4}
{{\cal D}_1{\cal D}_2}
+\frac{\theta(x_2-y_2)}{(x_2-y_2)^2}
\biggl( \frac{{\cal D}_2+{\cal D}_4}{{\cal D}_3{\cal D}_4} 
+ \frac{{\cal D}_2-{\cal D}_4}{{\cal D}_5{\cal D}_6}
\biggr)
\biggr]
\nonumber\\
&=&-8\pi\al_s C_F(P_A+iQ_A)\biggl[\frac{1}{(y_2-x_2){\cal D}_1{\cal D}_2}\biggr]
-8\pi\al_s C_F\biggl[\frac{2}{(y_2-x_2)^2{\cal D}_2}\biggr],
\eea
and
\bea\label{A6}
\biggl[\sum_iT_{B_i}^{(\up\down\to\up\down)}\biggr]_{\Delta=0}
&=&-8\pi\al_s C_F\biggl[
\frac{\theta(y_2-x_2)(P_{B_1} + iQ_B)}{(y_2-x_2){\cal D}_7{\cal D}_8}
+\frac{\theta(y_2-x_2)(P_{B3}+iQ_B)}{(y_2-x_2){\cal D}_{11}{\cal D}_{12}}
+\frac{\theta(x_2-y_2)(P_{B2} + iQ_B)}{(x_2-y_2){\cal D}_9{\cal D}_{10}}
\biggr]
\nonumber\\
&=&-8\pi\al_s C_F(P_B+iQ_B)\biggl[
\frac{\theta(y_2-x_2)}{(y_2-x_2)}
\biggl(\frac{1}{{\cal D}_7{\cal D}_8}+\frac{1}{{\cal D}_{11}{\cal D}_{12}}
\biggr)
+\frac{\theta(x_2-y_2)}{(x_2-y_2){\cal D}_9{\cal D}_{10}}
\biggr]
\nonumber\\
&&-16\pi\al_s C_F\biggl[
\frac{\theta(y_2-x_2)}{(y_2-x_2)^2}
\biggl(
\frac{{\cal D}_9+{\cal D}_7}{{\cal D}_7{\cal D}_8} 
+ \frac{{\cal D}_9-{\cal D}_7}{{\cal D}_{11}{\cal D}_{12}}
\biggr)
+\frac{\theta(x_2-y_2)}{(x_2-y_2)^2}\frac{{\cal D}_9+{\cal D}_7}
{{\cal D}_9{\cal D}_{10}}
\biggr]
\nonumber\\
&=&-8\pi\al_s C_F(P_B+iQ_B)\biggl[\frac{1}{(y_2-x_2){\cal D}_2{\cal D}_8}\biggr]
-8\pi\al_s C_F\biggl[\frac{2}{(y_2-x_2)^2{\cal D}_5}\biggr],
\eea
\end{widetext}
where the first terms in Eqs.~(\ref{A5}) and~(\ref{A6}) proportional
to $1/(y_2-x_2)$ and the second terms proportional to $1/(y_2-x_2)^2$
are related with the Feynman gauge and the LF gauge parts, respectively.
By adding all six LF time-ordered diagrams, we obtain
\bea\label{A7}
T_{H}^{(\up\down\to\up\down)}
&=&
\sum_i\biggl[T_{A_i}^{(\up\down\to\up\down)}
+ T_{B_i}^{(\up\down\to\up\down)}\biggr]_{\Delta=0}
\nonumber\\
&=&-\frac{8\pi\al_s C_F}{(y_2-x_2)}
\biggl[\frac{(P_A+iQ)}{{\cal D}_1{\cal D}_2}+
\frac{(P_B+iQ_B)}{{\cal D}_2{\cal D}_8}\biggr],
\nonumber\\
\eea
i.e. the singular LF gauge parts cancel each other and only finite Feynman 
gauge parts contribute to the amplitude. Similarly, we obtain other helicity
contributions to the hard scattering amplitude as shown 
in Tables~\ref{t4} and~\ref{t5}.

\section{Hard scattering amplitude combined with
Relativistic Spin-orbit wave function}
In this appendix B, we list the leading and subleading helicity 
contributions to the hard scattering amplitude combined with the 
relativistic spin-orbit wave function, where the subleading helicity
contributions show up as next-to-leading order in transverse momenta. 
That is, the subleading helicity contributions vanish at leading twist.

We first consider the relativistic spin-orbit wave functions for pseudoscalar
and vector(with transverse polarization $\ep=+1$)
mesons given by Eqs.~(\ref{R00}) and~(\ref{R11}), respectively.
Besides the leading helicity(in tranverse momenta) 
contributions coming from two $\Delta H=1$ contributions(i.e.
$\up\down\to\up\up$ and $\down\up\to\up\up$), the subleading helicity
contributions are as follows.

(1) $\Delta H=0$ contributions:
\bea\label{B1}
{\cal R}^{11\dagger}_{\up\down}{\cal R}^{00}_{\up\down}
&=&\frac{m\sqrt{2}}{C_xC_y}
\biggl[
\frac{y_1M_{0y}+m}{M_{0y}+2m}
\biggr]l^L
= -{\cal R}^{11\dagger}_{\up\down}{\cal R}^{00}_{\down\up},
\nonumber\\
{\cal R}^{11\dagger}_{\down\up}{\cal R}^{00}_{\down\up}
&=&\frac{m\sqrt{2}}{C_xC_y}
\biggl[\frac{y_2M_{0y}+m}{M_{0y}+ 2m}\biggr]l^L
=-{\cal R}^{11\dagger}_{\down\up}{\cal R}^{00}_{\up\down},
\nonumber\\
{\cal R}^{11\dagger}_{\up\up}{\cal R}^{00}_{\up\up}
&=&-\frac{\sqrt{2}}{C_xC_y}
\biggl[ m + \frac{{\bf l}^2_\perp}{M_{0y}+ 2m}
\biggr]k^L
\nonumber\\
{\cal R}^{11\dagger}_{\down\down}{\cal R}^{00}_{\down\down}
&=&\frac{\sqrt{2}}{C_xC_y}
\biggl[ \frac{1} {M_{0y}+2m} \biggr]k^R(l^L)^2.
\eea
(2) $\Delta H=1$ contributions:
\bea\label{B2}
{\cal R}^{11\dagger}_{\down\down}{\cal R}^{00}_{\down\up}
&=&\frac{m\sqrt{2}}{C_xC_y}
\biggl[ \frac{1}{M_{0y}+2m} \biggr](l^L)^2
= -{\cal R}^{11\dagger}_{\down\down}{\cal R}^{00}_{\up\down},
\nonumber\\
{\cal R}^{11\dagger}_{\up\down}{\cal R}^{00}_{\up\up}
&=&-\frac{\sqrt{2}}{C_xC_y}
\biggl[ \frac{y_1 M_{0y} + m}{M_{0y}+2m}
\biggr]l^Lk^L,
\nonumber\\
{\cal R}^{11\dagger}_{\up\down}{\cal R}^{00}_{\down\down}
&=&-\frac{\sqrt{2}}{C_xC_y}
\biggl[ \frac{y_1 M_{0y} + m}{M_{0y}+2m}
\biggr]l^Lk^R,
\nonumber\\
{\cal R}^{11\dagger}_{\down\up}{\cal R}^{00}_{\up\up}
&=&\frac{\sqrt{2}}{C_xC_y}
\biggl[ \frac{y_2 M_{0y} + m}{M_{0y}+2m}
\biggr]l^Lk^L,
\nonumber\\
{\cal R}^{11\dagger}_{\down\up}{\cal R}^{00}_{\down\down}
&=&\frac{\sqrt{2}}{C_xC_y}
\biggl[ \frac{y_2 M_{0y} + m}{M_{0y}+2m}
\biggr]l^Lk^R.
\eea
where $C_x=\sqrt{2x_1x_2}M_{0x}$ and $C_y=\sqrt{2y_1y_2}M_{0y}$. 
Since the hard scattering
amplitudes vanish for $\Delta H=2$ cases,
we do not consider them here. 

Next, we obtain the hard scattering amplitude combined with the spin-orbit
wave function.

(1) $\Delta H=0$ contributions:
\bea\label{B3}
{\cal T}^{R}_H(\up\down\to\up\down)
&=&\frac{\sqrt{2}}{q^L}
{\cal R}^{11\dagger}_{\up\down}T^{(\up\down\to\up\down)}_H
{\cal R}^{00}_{\up\down}
\nonumber\\
&=&\frac{16\pi\al_s C_F m}{(y_2-x_2)C_xC_y q^2}
\biggl[ \frac{y_1M_{0y}+m}{M_{0y}+2m}\biggr]
\nonumber\\
&&\times(l^L q^R)
\biggl[ \frac{P_A+iQ_A}{{\cal D}_1{\cal D}_2} 
+ \frac{P_B+iQ_B}{{\cal D}_2{\cal D}_8} \biggr],
\nonumber\\
\eea
\bea\label{B4}
{\cal T}^{R}_H(\down\up\to\down\up)
&=&\frac{\sqrt{2}}{q^L}
{\cal R}^{11\dagger}_{\down\up}T^{(\down\up\to\down\up)}_H
{\cal R}^{00}_{\down\up}
\nonumber\\
&=&\frac{16\pi\al_s C_F m}{(y_2-x_2)C_xC_y q^2}
\biggl[ \frac{y_2M_{0y}+m}{M_{0y}+2m} \biggr]
\nonumber\\
&&\times(l^L q^R)
\biggl[ \frac{P_A-iQ_A}{{\cal D}_1{\cal D}_2} 
+ \frac{P_B-iQ_B}{{\cal D}_2{\cal D}_8} \biggr],
\nonumber\\
\eea
\bea\label{B5}
{\cal T}^{R}_H(\up\up\to\up\up)
&=&\frac{\sqrt{2}}{q^L}
{\cal R}^{11\dagger}_{\up\up}T^{(\up\up\to\up\up)}_H
{\cal R}^{00}_{\up\up}
\nonumber\\
&=&\frac{-16\pi\al_s C_F}{(y_2-x_2)C_xC_y q^2}
\biggl[ m + \frac{{\bf l}^2_\perp}{M_{0y}+2m} \biggr]
\nonumber\\
&&\times(k^L q^R)
\biggl[ \frac{F_A+iG_A}{{\cal D}_1{\cal D}_2} 
+ \frac{F_B+iG_B}{{\cal D}_2{\cal D}_8} \biggr],
\nonumber\\
\eea
\bea\label{B6}
{\cal T}^{R}_H(\down\down\to\down\down)
&=&\frac{\sqrt{2}}{q^L}
{\cal R}^{11\dagger}_{\down\down}T^{(\down\down\to\down\down)}_H
{\cal R}^{00}_{\down\down}
\nonumber\\
&=&\frac{16\pi\al_s C_F}{(y_2-x_2)C_xC_y q^2}
\biggl[ \frac{1}{M_{0y}+2m} \biggr]
\nonumber\\
&&\times(k^Rl^Ll^L q^R)
\biggl[ \frac{F_A-iG_A}{{\cal D}_1{\cal D}_2} 
+ \frac{F_B-iG_B}{{\cal D}_2{\cal D}_8} \biggr],
\nonumber\\
\eea
\bea\label{B7}
{\cal T}^{R}_H(\up\down\to\down\up)
&=&\frac{\sqrt{2}}{q^L}
{\cal R}^{11\dagger}_{\down\up}T^{(\up\down\to\down\up)}_H
{\cal R}^{00}_{\up\down}
\nonumber\\
&=&\frac{32\pi\al_s C_F m^3 (y_1-x_1)}{x_1x_2y_1y_2C_xC_y q^2}
\biggl[ \frac{y_2M_{0y}+m}{M_{0y}+2m} \biggr]
\nonumber\\
&&\times(l^L q^R)
\biggl[ \frac{1}{{\cal D}_1{\cal D}_2} 
+ \frac{1}{{\cal D}_2{\cal D}_8} \biggr],
\eea
\bea\label{B8}
{\cal T}^{R}_H(\down\up\to\up\down)
&=&\frac{\sqrt{2}}{q^L}
{\cal R}^{11\dagger}_{\up\down}T^{(\down\up\to\up\down)}_H
{\cal R}^{00}_{\down\up}
\nonumber\\
&=&\frac{32\pi\al_s C_F m^3 (y_1-x_1)}{x_1x_2y_1y_2C_xC_y q^2}
\biggl[ \frac{y_1M_{0y}+m}{M_{0y}+2m} \biggr]
\nonumber\\
&&\times(l^L q^R)
\biggl[ \frac{1}{{\cal D}_1{\cal D}_2} 
+ \frac{1}{{\cal D}_2{\cal D}_8} \biggr].
\eea
(2) $\Delta H=1$ contributions:
\bea\label{B9}
{\cal T}^{R}_H(\up\down\to\down\down)
&=&\frac{\sqrt{2}}{q^L}
{\cal R}^{11\dagger}_{\down\down}T^{(\up\down\to\down\down)}_H
{\cal R}^{00}_{\up\down}
\nonumber\\
&=&\frac{-32\pi\al_s C_F m^2}{x_1y_1y_2(y_2-x_2)C_xC_y q^2}
\biggl[ \frac{1}{M_{0y}+2m} \biggr]
\nonumber\\
&&\times(l^L q^R)
\biggl[ \frac{x_2{\bf l}^2_\perp 
- y_2l^Lk^R - x_2y_2l^Lq^R}{{\cal D}_1{\cal D}_2} 
\nonumber\\
&&+ \frac{x_2{\bf l}^2_\perp 
- y_2l^Lk^R - x_2y_2l^Lq^R}{{\cal D}_2{\cal D}_8} \biggr],
\eea
\bea\label{B10}
{\cal T}^{R}_H(\down\up\to\down\down)
&=&\frac{\sqrt{2}}{q^L}
{\cal R}^{11\dagger}_{\down\down}T^{(\down\up\to\down\down)}_H
{\cal R}^{00}_{\down\up}
\nonumber\\
&=&\frac{-32\pi\al_s C_F m^2}{x_2y_1y_2(y_2-x_2)C_xC_y q^2}
\biggl[ \frac{1}{M_{0y}+2m} \biggr]
\nonumber\\
&&\times(l^L q^R)
\biggl[ \frac{x_1{\bf l}^2_\perp-y_1l^Lk^R-x_2y_1l^Lq^R}
{{\cal D}_1{\cal D}_2} 
\nonumber\\
&&+ \frac{x_1{\bf l}^2_\perp-y_1l^Lk^R-x_1y_2l^Lq^R}
{{\cal D}_2{\cal D}_8} \biggr],
\eea
\bea\label{B11}
{\cal T}^{R}_H(\up\up\to\up\down)
&=&\frac{\sqrt{2}}{q^L}
{\cal R}^{11\dagger}_{\up\down}T^{(\up\up\to\up\down)}_H
{\cal R}^{00}_{\up\up}
\nonumber\\
&=&\frac{32\pi\al_s C_F m}{x_1x_2y_2(y_2-x_2)C_xC_y q^2}
\biggl[ \frac{y_1M_{0y}+m}{M_{0y}+2m} \biggr]
\nonumber\\
&&\times(l^L q^R)
\biggl[ \frac{x_1k^Ll^R - y_1{\bf k}^2_\perp 
- x_2y_1k^Lq^R}{{\cal D}_1{\cal D}_2}
\nonumber\\
&&+ \frac{x_1k^Ll^R - y_1{\bf k}^2_\perp 
- x_1y_2k^Lq^R}{{\cal D}_2{\cal D}_8} \biggr],
\eea
\bea\label{B12}
{\cal T}^{R}_H(\down\down\to\up\down)
&=&\frac{\sqrt{2}}{q^L}
{\cal R}^{11\dagger}_{\up\down}T^{(\down\down\to\up\down)}_H
{\cal R}^{00}_{\down\down}
\nonumber\\
&=&\frac{32\pi\al_s C_F m}{x_1x_2y_1(y_2-x_2)C_xC_y q^2}
\biggl[ \frac{y_1M_{0y}+m}{M_{0y}+2m} \biggr]
\nonumber\\
&&\times(l^L q^R)
\biggl[ \frac{x_2l^Lk^R - y_2{\bf k}^2_\perp-x_2y_2q^Lk^R}
{{\cal D}_1{\cal D}_2}
\nonumber\\
&&+ \frac{x_2l^Lk^R - y_2{\bf k}^2_\perp-x_2y_2q^Lk^R}
{{\cal D}_2{\cal D}_8} \biggr],
\eea
\bea\label{B13}
{\cal T}^{R}_H(\up\up\to\down\up)
&=&\frac{\sqrt{2}}{q^L}
{\cal R}^{11\dagger}_{\down\up}T^{(\up\up\to\down\up)}_H
{\cal R}^{00}_{\up\up}
\nonumber\\
&=&\frac{32\pi\al_s C_F m}{x_1x_2y_1(y_2-x_2)C_xC_y q^2}
\biggl[ \frac{y_2M_{0y}+m}{M_{0y}+2m} \biggr]
\nonumber\\
&&\times(l^L q^R)
\biggl[ \frac{x_2k^Ll^R - y_2{\bf k}^2_\perp 
- x_2y_2k^Lq^R}{{\cal D}_1{\cal D}_2}
\nonumber\\
&&+ \frac{x_2k^Ll^R - y_2{\bf k}^2_\perp 
- x_2y_2k^Lq^R}{{\cal D}_2{\cal D}_8} \biggr],
\eea
\bea\label{B14}
{\cal T}^{R}_H(\down\down\to\down\up)
&=&\frac{\sqrt{2}}{q^L}
{\cal R}^{11\dagger}_{\down\up}T^{(\down\down\to\down\up)}_H
{\cal R}^{00}_{\down\down}
\nonumber\\
&=&\frac{32\pi\al_s C_F m}{x_1x_2y_2(y_2-x_2)C_xC_y q^2}
\biggl[ \frac{y_2M_{0y}+m}{M_{0y}+2m} \biggr]
\nonumber\\
&&\times(l^L q^R)
\biggl[ \frac{x_1l^Lk^R-y_1{\bf k}^2_\perp-x_2y_1q^Lk^R}
{{\cal D}_1{\cal D}_2}
\nonumber\\
&&+ \frac{x_1l^Lk^R-y_1{\bf k}^2_\perp-x_1y_2q^Lk^R}
{{\cal D}_2{\cal D}_8} \biggr].
\eea


\begin{thebibliography}{99}
\bibitem{BJ85} S.J. Brodsky and C.-R. Ji, \Journal{\PRL}{55}{2257}{1985}.
\bibitem{BJ03} E. Braaten and J. Lee, \Journal{\PRD}{67}{054007}{2003};
{\bf 72}, 099901(E) (2005).
\bibitem{LHC03} K.-Y. Liu, Z.-G. He, and K.-T. Chao, 
\Journal{\PLB}{557}{45}{2003}.
\bibitem{HKQ03} K. Hagiwara, E. Kou, and C.-F. Qiao, 
\Journal{\PLB}{570}{39}{2003}.
\bibitem{Kis} V.V. Kiselev, Int. J. Mod. Phys. A {\bf 10}, 465 (1995).
\bibitem{VS} G.T. Bodwin, E. Braaten, and G.P. Lepage,
\Journal{\PRD}{51}{1125}{1995};{\bf 55}, 5853(E)(1997).
\bibitem{Belle} K. Abe {\em et al.}(Belle Collaboration),
\Journal{\PRL}{89}{142001}{2002}; \Journal{\PRD}{70}{071102}{2004}.
\bibitem{Babar}B. Aubert {\em et al.}(Babar Collaboration),
\Journal{\PRD}{72}{031101}{2005}.
\bibitem{BC} A.E. Bondar and V.L. Chernyak, \Journal{\PLB}{612}{215}{2005}.
\bibitem{Ma} J.P. Ma and Z.G. Si, \Journal{\PRD}{70}{074007}{2004}.
\bibitem{BLL05} V.V. Braguta, A.K. Likhoded, and A.V. Luchinsky,
\Journal{\PRD}{72}{074019}{2005}.
\bibitem{Huang} T. Huang and F. Zuo, arXiv:hep-ph/0702147v2.
\bibitem{JP} C.-R. Ji and A. Pang, \Journal{\PRD}{55}{1253}{1997}.
\bibitem{CJD} H.-M. Choi and C.-R. Ji, \Journal{\PRD}{73}{114020}{2006}.
\bibitem{CJ1} H.-M. Choi and C.-R. Ji, \Journal{\PRD}{59}{074015}{1999}.
\bibitem{CJ2} H.-M. Choi and C.-R. Ji, \Journal{\PLB}{460}{461}{1999}.
\bibitem{EM} D. Ebert and A.P. Martynenko, \Journal{\PRD}{74}{054008}{2006}.
\bibitem{ZGC} Y.-J. Zhang, Y.-J. Gao, and K.T. Chao,
\Journal{\PRL}{96}{092001}{2006}.
\bibitem{BL} G. P. Lepage and S. J. Brodsky, \Journal{\PRD}{22}{2157}{1980};
S.J. Brodsky, T. Huang, and G.P. Lepage, in {\em Particles and Fields-2},
Proceedings of the Banff Summer Institute, Banff, Alberta, 1981, edited by
A.Z.Capri and A.N. Kamal(Plenum, New York, 1983), p. 143.
\bibitem{CJ_DA} H.-M. Choi and C.-R. Ji, \Journal{\PRD}{75}{034019}{2007}.
\bibitem{CJadd} H.-M. Choi and C.-R. Ji, \Journal{\PRD}{72}{013004}{2005}.
\bibitem{JPS} C.-R. Ji, A. Pang, and A. Szczepaniak,
\Journal{\PRD}{52}{4038}{1995}.
\bibitem{HWW} T. Huang, X.-G. Wu, and X.-H. Wu,
\Journal{\PRD}{70}{053007}{2004}.
\bibitem{GI} S. Godfrey and N. Isgur, \Journal{\PRD}{32}{189}{1985};
S. Godfrey, \Journal{\PRD}{33}{1391}{1986}.
\bibitem{ISGW2} D. Scora and N. Isgur, \Journal{\PRD}{52}{2783}{1995}.
\bibitem{Cor} E. Eichten, K. Gottfried, T. Kinoshita, K.D. Lane
and T.M. Yan,
\Journal{\PRD}{17}{3090}{1978}[Erratum-ibid. D {\bf 21}, 313 (1980)].
\bibitem{BKL} G.T. Bodwin, D. Kang and J. Lee,
\Journal{\PRD}{74}{114028}{2006}.
\bibitem{Bali} G.S. Bali, Phys. Rept. {\bf 343}, 1(2001).
\bibitem{CLEO01} K.W. Edwards {\em et al.}, CLEO Collaboration,
\Journal{\PRL}{86}{30}{2001}.
\bibitem{PDG06} W.-M. Yao {\em et al.}(Particle Data Group),
\Journal{\JPG}{33}{1}{2006}.
\bibitem{BLL} V.V. Braguta, A.K. Likhoded, and A.V. Luchinsky,
\Journal{\PLB}{646}{80}{2007}.
\bibitem{Bra} V.V. Braguta, \Journal{\PRD}{75}{094016}{2007}.
\bibitem{BT} W. Buchmuller and S.H.H. Tye, \Journal{\PRD}{24}{132}{1981}.
\bibitem{BKL2} G.T Bodwin, D. Kang and J. Lee,
\Journal{\PRD}{74}{014014}{2006}.
\bibitem{JMY} X.D. Ji, J.P. Ma and F. Yuan, 
\Journal{\PRL}{90}{241601}{2003}.
\bibitem{BF} S.J. Brodsky and G.R. Farrar, \Journal{\PRL}{31}{1153}{1973};
\Journal{\PRD}{11}{1309}{1975}.
\bibitem{MMT} V.A. Matveev, R.M. Muradian and A.N. Tavkhelidze, 
Nuovo Cim. Lett. {\bf 7}, 719 (1973).
\bibitem{CJ} C.E. Carlson and C.-R. Ji,
\Journal{\PRD}{67}{116002}{2003}.
\bibitem{BJ} B.L.G. Bakker and C.-R. Ji, 
\Journal{\PRD}{65}{073002}{2002}.
\end{thebibliography}
\end{document}